\documentclass[aps,prd,twocolumn,superscriptaddress,groupedaddress, nofootinbib]{revtex4}  
\usepackage{graphicx}
\usepackage{dcolumn}
\usepackage{bm}
\usepackage{hyperref}

\usepackage{epsfig,graphics,colordvi}
\usepackage{amsmath}
\usepackage{amssymb}
\usepackage{mciteplus}

\usepackage{graphicx}
\usepackage[english]{babel}
\usepackage{caption}
\usepackage{sidecap}
\usepackage{multirow}

\usepackage{mathrsfs}

\usepackage{color}
\usepackage[usenames,dvipsnames]{xcolor}
\usepackage{bbold}

\newcommand{\tr}{\operatorname{tr}}
\newcommand{\te}{\text}
\newcommand{\nn}{\nonumber}

\newcommand{\eps}{\varepsilon}
\newcommand{\etapr}{\eta\hspace{0.05em}'}
\newcommand{\df}{\!\text{d}^4}

\newcommand{\Lag}{\mathscr{L}}
\newcommand{\order}{\mathcal{O}}
\newcommand{\chpt}{{\chi\te{PT}}}
\newcommand{\bra}{\left\langle}
\newcommand{\ket}{\right\rangle}

\newcommand{\Dvec}{D_{\te{vec}}}
\newcommand{\Dmix}{D_{\te{mix}}}
\newcommand{\Dch}{D_{\te{pseudo}}}
\newcommand{\Lvec}{\Lag_{\te{vec}}}
\newcommand{\Lfree}{\Lag_{\te{free}}}
\newcommand{\Llin}{\Lag_{\te{lin}}}
\newcommand{\Lch}{\Lag_\chpt}
\newcommand{\Vt}{\widetilde{V}}
\newcommand{\Um}{\mathcal{U}}
\newcommand{\Umb}{\bar{\mathcal{U}}}
\newcommand{\Ub}{\bar{U}}
\newcommand{\ub}{\bar{u}}

\newcommand{\Mp}{\mathcal{M}_{\! P}}
\newcommand{\mv}{m_V}

\newcommand{\sd}{{\cdot}}
\newcommand{\adb}{\allowdisplaybreaks}

\DeclareMathOperator{\GammaF}{\operatorname{\Gamma}}

\newcommand{\Apr}{\mathcal{A}}
\newcommand{\Vpr}{\mathcal{V}}

\begin{document}

\title{Renormalisation of the low-energy constants of chiral perturbation theory from loops with dynamical vector mesons}

\author{Carla Terschl\"usen and Stefan Leupold}  

\affiliation{Institutionen f\"or fysik och astronomi, Uppsala Universitet, Box 516, 75120 Uppsala, Sweden}


\begin{abstract}
Starting from a relativistic Lagrangian for pseudoscalar Goldstone bosons and vector mesons in the antisymmetric tensor representation, a one-loop calculation is performed to pin down the divergent structures that appear for the effective low-energy action at chiral orders $Q^2$ and $Q^4$. The corresponding renormalisation-scale dependences of all low-energy constants up to chiral order $Q^4$ are determined. Calculations are carried out for both the pseudoscalar octet and the pseudoscalar nonet, the latter in the framework of chiral perturbation theory in the limit of a large number of colours.
\end{abstract}
%
%
\maketitle


\section{Introduction and summary} \label{sec:Intro}

\subsection{Scale separation} \label{subsec:scale-separation}
Chiral perturbation theory ($\chpt$) \cite{Weinberg:1978kz, Gasser:1983yg, Gasser:1984gg, Scherer:2002tk,Scherer-Schindler}, the low-energy incarnation of the non-perturbative aspects of the standard model of particle physics, is based on a separation of scales. This separation allows for systematic power counting and qualifies $\chpt$ as an effective field theory. The dynamical (low-energy/soft) scale is provided by the masses of the lowest pseudoscalar multiplet, the Goldstone bosons. Their smallness is caused by the smallness of the current quark masses of the lightest (two or three) quark flavours. To be more specific, the spontaneous breaking of chiral symmetry demands the appearance of massless pseudoscalar Goldstone bosons. The explicit breaking of chiral symmetry by the current quark masses induces non-vanishing masses for these pseudoscalars. But the masses are small as compared to typical hadronic scales. The latter are related to the scale $\Lambda_{\rm QCD}$ where the strong interaction really becomes strong, which in turn is caused by the scale anomaly of the theory \cite{Pascual:1984zb,Bailin-Love}. 

Coming back to the scale separation, the static (high-energy/hard) scale is given by the typical hadronic scales. Conceptually it is useful to distinguish between different high-energy scales \cite{Aydemir:2012nz}. The ``external'' high-energy scale is the energy where neglected degrees of freedom become important. For chiral perturbation theory this scale is at least given by the vector-meson mass $m_V \approx 0.77\,$GeV of the $\omega$ and $\rho$ mesons \cite{Ecker:1988te,Donoghue:1988ed}, if not by the mass of the somewhat lighter $\sigma$ meson \cite{PDG2014}. The ``intrinsic'' high-energy scale is given by the energy where loops become as important as tree-level diagrams. For chiral perturbation theory this scale is roughly at $4\pi F_\pi \approx 1.2\,$GeV. 

Conceptually the scale separation provides a clear-cut power counting scheme if the momenta of the considered processes are on the order of the Goldstone-boson masses. Expansions are carried out around the formal limit where the considered momenta vanish along with the Goldstone-boson masses. The latter takes place in the chiral limit. When it comes to the real world where the current quark masses do not vanish, it is clear that the convergence of the expansions is the better the smaller the dynamical scale is relative to the static scale. For the two lightest quark flavours there is a large scale separation between the pion mass and corresponding momenta at close-to-threshold processes on the one side and the typical hadronic scales mentioned above on the other side. Including strangeness, however, with a kaon mass (dynamical scale) of about 500 MeV and a $K^*$ mass (degree of freedom that is integrated out) of about 900 MeV, the scales already move significantly together \cite{PDG2014}.

Another formally clear-cut power counting scheme, where however the numerical values for the dynamical and the static scale gets even more intertwined, is $\chpt$ for a large number of colours, $N_c$ \cite{'tHooft:1973jz, Witten:1979kh}. In the combined $N_c \to \infty$ and chiral limit the mass of the $\eta'$ meson vanishes \cite{Witten:1979vv, Veneziano:1979ec}. The pseudoscalar octet is enlarged to a nonet. Systematic expansions in powers of $1/N_c$, masses of the nonet states and momenta become possible \cite{Kaiser:2000gs}. Schemes based on $\chpt$ and the large-$N_c$ expansion \cite{Dashen:1993jt, Pascalutsa:2006up, Ledwig:2014rfa} lead to many phenomenologically appealing results in spite of the fact that in the real world the mass of the $\eta'$ is not at all lower than the masses of mesonic resonances like the vector mesons. In the large-$N_c$ limit one has the ordering
\begin{eqnarray}
  \label{eq:orderln}
  m_{\eta'}^2 \ll m_V^2 \ll (4 \pi F_\pi)^2  \,.
\end{eqnarray}
The first quantity scales like $1/N_c$ \cite{Witten:1979vv, Veneziano:1979ec}. The mass of a typical mesonic resonance, here the vector-meson mass $m_V$, scales like $1/N_c^0$. Finally the scale where loops become as important as tree-level processes, $(4 \pi F_\pi)^2$, scales like $N_c$. In the real world (\ref{eq:orderln}) is contrasted by
\begin{eqnarray}
  m_V^2 < m_{\eta'}^2 < (4 \pi F_\pi)^2  \,. 
  \label{eq:orderlnpr}
\end{eqnarray}
Nonetheless the large-$N_c$ approximation provides many insights in the dynamics of hadrons \cite{'tHooft:1973jz, Witten:1979kh, Witten:1979vv, Veneziano:1979ec, Dashen:1993jt, Kaiser:2000gs, Pascalutsa:2006up, Ledwig:2014rfa}.

\subsection{Excursion to baryons} \label{subsec:baryons}
The previous discussions provide a motivation why one might want to include additional degrees of freedom on top of the Goldstone bosons. Before addressing the central aspect of this work, the inclusion of vector mesons, it is illuminating to discuss a better established case where additional degrees of freedom have been included in the framework of chiral perturbation theory, namely the case of baryons \cite{Gasser:1987rb}. As always one has to distinguish the two- \cite{Hemmert:1997ye,Fettes:1998ud,Pascalutsa:2005vq,Pascalutsa:2007wb,Scherer-Schindler, Alarcon:2012kn} and three-flavour \cite{Jenkins:1990jv,Jenkins:1991es,Kubis:2000aa,Lutz:2001yb,Frink:2006hx,Geng:2008mf,Ledwig:2014rfa} case and it should be clear that the scale separation and therefore the convergence properties are better for the two-flavour case. But, in addition, it matters whether the scheme treats the baryons relativistically \cite{Becher:1999he,Gegelia:1999gf,Kubis:2000aa,Lutz:2001yb,Fuchs:2003qc,Pascalutsa:2005vq,Pascalutsa:2007wb,Frink:2006hx,Geng:2008mf,Scherer-Schindler,Ledwig:2014rfa,Alarcon:2012kn} or non-relativistically \cite{Jenkins:1990jv,Jenkins:1991es,Hemmert:1997ye,Fettes:1998ud} and whether \cite{Jenkins:1990jv,Jenkins:1991es,Hemmert:1997ye,Lutz:2001yb,Pascalutsa:2005vq,Pascalutsa:2007wb,Ledwig:2014rfa,Alarcon:2012kn} or not \cite{Fettes:1998ud,Kubis:2000aa,Frink:2006hx,Geng:2008mf,Scherer-Schindler} the decuplet (for two flavours: the Delta iso-quartet) is included on top of the ground-state baryon octet (for two flavours: the nucleon iso-doublet). 

Before addressing these issues we should stress right away that the inclusion of baryonic degrees of freedom in a chiral effective field theory framework is conceptually much more straight forward than the inclusion of (non-Goldstone) mesonic degrees of freedom (meson resonances). Because of baryon number conservation a heavy (static) scale --- the baryon mass --- remains in the considered process from beginning to end. The small (dynamical) scales are then given by the masses of the Goldstone bosons, the three-momenta of the involved particles and the mass differences between the baryon states. In contrast, for a meson resonance one has to deal with the fact that this resonance can decay into Goldstone bosons. If one treats the resonance mass as a heavy (static) scale, like the baryon mass, then this implies that the momenta of the emerging Goldstone bosons cannot (all) be soft \cite{Bijnens:1997ni,Bijnens:1998di,Bruns:2004tj,Djukanovic:2014rua}. One suggestion to deal with this problem is the hadrogenesis conjecture \cite{Lutz:2001mi,Kolomeitsev:2003kt,Lutz:2003fm,Lutz:2008km, Terschlusen:2012xw,Danilkin:2011fz,Danilkin:2012ua} where a significant mass gap is proposed between the $J^P=0^-$, $1^-$, $\frac{1}{2}^+$, and $\frac{3}{2}^+$ ground states on the one hand and all other large-$N_c$ stable hadrons on the other hand. In this scheme the vector-meson mass constitutes a dynamical/soft scale. Consequently all Goldstone bosons emerging from vector-meson decays have soft momenta. The work presented here is fully compatible with the hadrogenesis conjecture, but is not restricted to it. In the present work and in \cite{MassPaper} we explore the quantitative impact of one-loop contributions with dynamical vector mesons on the low-energy effective action and on the properties of pseudoscalar mesons. Vector-meson masses and coupling constants are adopted from phenomenology. The formal power counting of the vector-meson mass will be of little concern as we will fully integrate out the vector mesons. We will come back to this point below after discussing the case of baryon $\chpt$.  

In spite of the conceptual difference between the inclusion of baryons or mesons we want to use the better established case of including baryonic degrees of freedom to discuss two issues relevant for both cases (meson and baryon): First, connected to the previous discussion around \eqref{eq:orderln}, \eqref{eq:orderlnpr}, the issue how well or not well separated the static and the dynamical scales actually are in practice. Second, the important technical issue how to deal with loops that contain non-Goldstone bosons. 

In the chiral limit one can find a momentum regime where only the ground-state baryons and the Goldstone bosons are active degrees of freedom. In reality, however, the mass difference between Delta and nucleon is not very large \cite{PDG2014}. In fact, in the combined chiral and large-$N_c$ limit (and ignoring electromagnetic effects) the nucleon and Delta become degenerate \cite{Dashen:1993jt}. Thus it might make sense to include the Deltas (and their flavour partners) as active degrees of freedom. Of course, this adds credits to the central theme of this work, the inclusion of additional degrees of freedom. 

If baryons are included in chiral perturbation theory, it turns out that the naive chiral power counting of loops is spoiled by the appearance of the additional static scale, the (average) baryon mass \cite{Gasser:1987rb}. This problem will not show up, if one treats the baryons non-relativistically (heavy-baryon chiral perturbation theory). In principle, all contributions from a non-relativistic expansion (Foldy–-Wouthuysen expansion) of relativistic interactions and propagators show up at appropriate orders in the chiral power counting. In reality, however, it turns out that often better results are obtained with a fully relativistic framework, see, \textit{e.g.}, \cite{Kubis:2000aa,Geng:2008mf,Scherer-Schindler}. If the convergence properties were excellent, this would not matter. In reality it does to some extent, even for the case of two flavours.  

In a relativistic setup there are in principle two possibilities how to deal with loop integrals: a) One splits up each integral in two parts, one that is in accordance with the chiral power counting and one that is not. The latter is then disregarded. We note in passing that there are several ways how to perform this splitting of integrals \cite{Becher:1999he,Gegelia:1999gf,Lutz:2001yb,Fuchs:2003qc}. The quality of convergence might depend on the way that one chooses \cite{Geng:2008mf}. The alternative, b) is to keep the integrals as they are. As a consequence the integrals do not only contribute at the chiral order that is formally assigned to them. Instead (polynomial parts of) the integrals contribute to lower, \textit{i.e.}, more important orders of the chiral expansion. Corresponding low-energy constants from these lower orders serve to renormalise the loops \cite{Gasser:1987rb}. 
This is the approach that we follow in the present work. 

To summarise the discussion of baryon chiral perturbation theory: The more separated the hard and soft scales are, the less it matters how one includes heavy degrees of freedom. But the closer the scales move to each other, the more problematic it might become to ignore the loops with additional degrees of freedom or parts of these loops. Consequently we will use in the present work a fully relativistic framework and identify explicitly the counter terms for the loop divergences irrespective of the formal chiral order of the loops and counter terms. 

\subsection{Inclusion of vector mesons} \label{subsec:incl-vector-meson}
While there is a clear gap between the masses of the lightest pseudoscalar mesons and the masses of other hadrons built from the lightest two quark flavours, the mass difference for the light vector mesons and the $\eta$ meson is not that big anymore. The $\etapr$ meson is even heavier than most of the vector mesons from the lowest-lying multiplet. All this concerns the physical masses. On the theory side there is one more situation where the dynamical and the static scale move closer together: Still until today a significant part of lattice-QCD calculations deals with too heavy ``light'' quark masses \cite{Gattringer:2010zz}. 
Therefore, it is valid to discuss if and, if yes, which hadrons should be included as additional degrees of freedom in an extended effective theory. The lightest non-Goldstone boson, the $\sigma$ meson is a notoriously complicated state; see, for instance, the discussion in \cite{PDG2014} on low-lying scalars. In addition, it is a very broad resonance. Thus its general impact might be limited. On the other hand, the low-lying vector mesons have both masses close to the Goldstone-boson masses and small widths. Thus, they are expected to be prominent in an effective theory including Goldstone bosons and other light mesons. 

As already mentioned, the inclusion of addional mesonic degrees of freedom in an effective theory is not free of complications and/or input assumptions. Concerning the scale separation one complication is caused by the fact that numerically the masses of the vector mesons are similar to the scale $4\pi F_\pi$ where loops become as important as tree-level contributions, see \eqref{eq:orderln}, \eqref{eq:orderlnpr}. Here a possible solution could come from the resummation of the numerically most important loop diagrams \cite{Lutz:2001yb,Danilkin:2011fz,Danilkin:2012ua,Lutz:2015lca}.

Another important issue is the representation dependence. In principle it should not matter for an effective theory whether vector degrees of freedom are represented, \textit{e.g.}, by ordinary vector fields, massive Yang-Mills fields, hidden gauge fields or antisymmetric tensor fields; see, for instance, the discussions in \cite{Ecker:1989yg, Bijnens:1995ii, Knecht:2001xc}. However, the explicit power counting, \textit{i.e.}, the classification of interaction terms and diagrams might change when changing the representation. 

In the present work we have a much more modest aim than setting up and/or checking the validity of a power counting scheme for vector mesons. Here and in the follow-up work \cite{MassPaper} we will check the quantitative influence of one-loop contributions with dynamical vector mesons. We have chosen the antisymmetric tensor representation based on its phenomenological success; see, \textit{e.g.}, \cite{Gasser:1983yg, Ecker:1988te, Terschluesen:2010ik, Terschlusen:2013iqa}. The present work should be understood as a feasibility study for one-loop calculations with vector mesons in the antisymmetric tensor representation. In addition, we intend to scrutinise the effective-field-theory assumption that at low --- but practically relevant! --- energies the influence of vector mesons can fully be accounted for by the low-energy constants of the chiral Lagrangian. Starting out from a Lagrangian with vector mesons one will obtain a non-local effective action if one integrates out the vector mesons and the fluctuations in the pseudoscalar fields. The local part of this effective action, \textit{i.e.}, the polynomial terms can be matched by an adjustment of the low-energy constants. The non-local part, related to the logarithms emerging from the loop integrals, can only be matched, if it is further Taylor expanded. However, if this part is numerically significant, the Taylor expansion might not converge very well and jeopardise in that way the convergence of the chiral expansion. In the present work we address the cancellation of one-loop divergences by the counter terms provided in the form of the low-energy constants of $\chpt$. Equipped with the knowledge about these local structures we will address in the follow-up work \cite{MassPaper} the possible importance of the non-local logarithmic structures.

As already discussed, the inclusion of additional (mesonic) degrees of freedom in $\chpt$ is representation dependent. Vector mesons can be described as vectors or antisymmetric tensors or can be included via a hidden local gauge mechanism \cite{Harada:2003jx}. As a glance of this representation dependence we compare in this article our final results to those obtained from a hidden local gauge mechanism \cite{Harada:2003jx}.  

Aiming at a systematic inclusion of vector mesons as active degrees of freedom in an effective-field-theory framework we perform in the present work a feasibility study concerning renormalisation aspects at the one-loop level. We focus on the full effective actions at chiral order $Q^2$ and $Q^4$ where the vector mesons have been completely integrated out. This approach is complementary to the explicit calculation of selected $n$-point functions as carried out, for instance, in \cite{Kampf:2009jh} for vector-meson properties or in \cite{Pich:2008jm, SanzCillero:2009pt, Pich:2010sm, Guo:2014yva} for some low-energy constants of $\chpt$. Note that in the latter works not only vector mesons have been considered and also additional assumptions about the high-energy behaviour \cite{Cirigliano:2006hb} of resonance Lagrangians have been made there. We are aiming at the construction of a low-energy theory for the lowest-lying (vector-meson) resonances and do not claim that our theory is valid a high energies. Therefore, it is not possible to compare the divergences calculated in \cite{Pich:2008jm, SanzCillero:2009pt, Pich:2010sm, Guo:2014yva} with the results obtained within this article.

In the present work, we determine the infinity structure and the corresponding renormalisation-scale dependence of all low-energy constants up to chiral order $Q^4$ that is needed to compensate the corresponding effects from loops that include vector mesons. The found scale dependence should be qualitatively interpreted in the following way: The finite parts of the loops with vector mesons depend on the masses of vector and pseudoscalar mesons, on the external momenta, and on the renormalisation scale. For observables, (only) the scale dependence is compensated by the one of the low-energy constants. What is particularly interesting for observables is the impact of loops with vector mesons on the momentum dependence. Concerning results of lattice calculations also the impact on the quark-mass dependence is of interest. Based on dimensional arguments, it can be expected that at least part of the $\log \mu^2$ dependence which we uncover in the present work comes along with a $\log s$ and/or $\log m_P^2$ dependence of observables. Here, $\mu$ denotes the renormalisation scale, $s$ the square of a generic external momentum, and $m_P$ the mass of a pseudoscalar Goldstone boson. Detailed studies of these dependences of observables are delegated to future works, where one is already in progress \cite{MassPaper}. 

We concentrate in the present work on the appearing infinities as defined by a slightly modified MS-bar scheme according to \cite{Gasser:1984gg}. Technically we use non-perturbative path-integral methods to keep the full chiral structure of the effective Lagrangian instead of just calculating loops for specific $n$-point functions. In contrast to $\chpt$ one-loop calculations as carried out in \cite{Gasser:1983yg, Gasser:1984gg}, a standard heat-kernel technique cannot be used for vector mesons represented by antisymmetric tensor fields since these fields contain frozen, non-propagating degrees of freedom which have a different short-distance behaviour than the active, propagating degrees of freedom. This is an unfortunate finding because the standard heat-kernel technique keeps in every step the full chiral structure of the effective action and brings along recursive relations which simplify and systematise the calculations when proceeding from one chiral order to the next. We regard it as illuminating to devote a subsection to the discussion of this not-working technique before we present a formalism that does work and serves to isolate and classify the infinities of the loop calculations. The calculations are involved but a viable cross check emerges from the fact that the full chiral structure needs to be reconstructed in the end from several distinct expressions. In other words, the elegance of the heat-kernel technique of \cite{Gasser:1983yg, Gasser:1984gg} concerning the full chiral structure is lost, but technically a powerful cross check of the results has been gained. 

Given that the calculations are rather involved we have decided for this exploratory work that we limit the possible interaction terms between vector mesons and low-energy degrees of freedom. We only consider the (chiralised) three-flavour versions of the phenomenologically well known $\rho$-$2\pi$ and $\rho$-$v$ couplings where $v$ denotes an external vector source. Other interaction terms that might be relevant for a full effective theory of pseudoscalar and vector mesons are presented and discussed, \textit{e.g.}, in \cite{Terschlusen:2012xw,Terschlusen:2013iqa}.

The article is organised in the following way. In section \ref{sec:Gen} the building blocks and pertinent Lagrangians for pseudoscalar and vector mesons are introduced. It is discussed how one-loop contributions in this framework are calculated. Hereby, approaches for calculating one-loop contributions with vector mesons which are not applicable are discussed as well. The calculation itself is split up into two parts: At first, in section \ref{sec:Octet} we discuss one-loop contributions for $\chpt$ plus vector mesons and their influence on the low-energy constants of $\chpt$ for the case where one includes only the pseudoscalar Goldstone octet. Afterwards, the calculations are extended by including the $\eta$-singlet as well (section \ref{sec:etaprime}). All calculations are carried out up to (including) chiral order $Q^4$. In the last section, an outlook is given.

\section{General considerations} \label{sec:Gen}

In this section, techniques used to calculate the one-loop contributions of light vector mesons are introduced. We will also document (in subsections \ref{subsec:ExpInPhi} and \ref{subsec:Heat-kernel}) methods which were tested in order to calculate the one-loop contributions but turned out to be intractable.

Although in the classical sense effective theories are non-renormalisable, they can be renormalised order by order. In pure $\chpt$, a diagram containing $n$ loops is at least suppressed by order $Q^{2n}$ for a typical momentum $Q$ according to general power counting arguments \cite{Weinberg:1978kz,Gasser:1983yg, Gasser:1984gg}. To calculate diagrams up to $\order(Q^4)$ in pure $\chpt$, both tree-level diagrams based on the leading-order (LO) and next-to-leading-order (NLO) Lagrangian and loop diagrams based only on the LO Lagrangian have to be involved. In \cite{Gasser:1983yg, Gasser:1984gg}, the one-loop contributions to the effective action were calculated using the pure $\chpt$-Lagrangian describing pseudoscalar fields only. Based on the techniques used therein, one-loop contributions including light vector mesons are calculated in this article. Thereby, the calculations are first restricted to the pseudoscalar octet, the singlet is only included in section \ref{sec:etaprime}. These calculations are a feasibility check for loop calculations based on a Lagrangian that includes vector mesons (in the antisymmetric tensor representation).

In this article, the $\chpt$ power-counting scheme is used, \textit{i.e.}, both derivates and pseudoscalar masses are treated as soft while the vector masses are not, 
\begin{align}
	\partial_\mu, \, m_P \in \order(Q), \ \mv \in \order(1).
\end{align}
Thus the effective action will not contain vector mesons. They are fully integrated out.

In the following we will perform one-loop calculations based on the LO Lagrangian of $\chpt$ and on a vector-meson Lagrangian to be specified below. We focus in the present work on those infinities where the counter terms are provided by the low-energy constant of the $\chpt$ Lagrangians of LO, $\order(Q^2)$, and NLO, $\order(Q^4)$. Those Lagrangians are given by \cite{Gasser:1984gg} 
\begin{align}
	& \Lch^{\rm LO} = \frac{1}{4} F^2 \left\{ \bra D_\mu U^\dagger \, D^\mu U \ket + \bra \chi U^\dagger + \chi^\dagger U \ket \right\}  , \nn \adb \\
	& \Lch^{\rm NLO} = L_1 \bra D_\mu U^\dagger \, D^\mu U \ket^2 + L_2 \bra D_\mu U^\dagger \, D_\nu U \ket^2 \nn \adb \\
	& \phantom{\Lch^{\rm NLO} =} + L_3 \bra (D_\mu U^\dagger \, D^\mu U)^2 \ket \nn \adb \\
	& \phantom{\Lch^{\rm NLO} =} + L_4 \bra D_\mu U^\dagger \, D^\mu U \ket \bra \chi^\dagger U + \chi U^\dagger \ket \nn \adb \\
	&  \phantom{\Lch^{\rm NLO} =} + L_5 \bra (D_\mu U^\dagger \, D^\mu U) (\chi^\dagger U + U^\dagger \chi) \ket \nn \adb \\
	& \phantom{\Lch^{\rm NLO} =} + L_6 \bra \chi^\dagger U + \chi U^\dagger \ket^2 \nn \adb \\
	&  \phantom{\Lch^{\rm NLO} =} + L_7 \bra \chi^\dagger U - \chi U^\dagger \ket^2 + L_8 \bra \chi^\dagger U \chi^\dagger U + \chi U^\dagger \chi U^\dagger \ket \nn \adb \\
	&  \phantom{\Lch^{\rm NLO} =} - iL_9 \bra F_R^{\mu\nu} D_\mu U \,  D_\nu U^\dagger + F_L^{\mu\nu} D_\mu U^\dagger \, D_\nu U \ket \nn \adb \\
	&  \phantom{\Lch^{\rm NLO} =} + L_{10} \bra U^\dagger F_R^{\mu\nu} U F^L_{\mu\nu} \ket \nn \adb \\
	&  \phantom{\Lch^{\rm NLO} =} + H_1 \bra F^R_{\mu\nu} F_R^{\mu\nu} + F^L_{\mu\nu} F_L^{\mu\nu} \ket + H_2 \bra \chi^\dagger \chi \ket. \label{eq:ChPTLagr}
\end{align}
The matrix $U := \exp(i\Phi/F)$ describes the pseudoscalar fields with the octet matrix
\begin{align}
	\Phi = \begin{pmatrix} \pi^0 + \frac{1}{\sqrt{3}}\eta_8 & \sqrt{2} \pi^+ & \sqrt{2} K^+ \\
							\sqrt{2} \pi^- & -\pi^0 + \frac{1}{\sqrt{3}} \eta_8 & \sqrt{2} K^0 \\
							\sqrt{2} K^- & \sqrt{2} \bar{K}^0 & -\frac{2}{\sqrt{3}} \eta_8
			\end{pmatrix} \label{eq:Def-pseudo-octet}
\end{align}
while the external vector, axialvector, scalar and pseudoscalar sources $v_\mu$, $a_\mu$, $s$ and $p$ are included in $F^R_\mu := v_\mu + a_\mu$, $F^L_\mu := v_\mu - a_\mu$ and $\chi := 2 B_0 (s+i p)$. Throughout this work we ignore isospin breaking effects. Thus we use an averaged quark mass $m_q$ for up and down quarks. The strange-quark mass $m_s$ is kept distinct. If the external fields are switched off, $\chi = 2 B_0 \mathcal{M} := 2 B_0 \operatorname{diag}(m_q, m_q, m_s)$. Furthermore, $\bra A \ket := \tr (\!A)$ and\footnote{Note that the chirally covariant derivative $D_\mu$ is defined depending on the field it is acting on and acts differently on $U$, $U^\dagger$ and the vector field $V$.}
\begin{align}
	& D_\mu U := \partial_\mu U - i F^R_\mu U + i U F^L_\mu\,, \nn \adb \\
	& D_\mu U^\dagger := \partial_\mu U^\dagger + i U^\dagger F_R^\mu - i F_L^\mu U^\dagger \,, \nn \adb \\
	& F_{\mu\nu}^{R/L} := \partial_\mu F_\nu^{R/L} - \partial_\nu F_\mu^{R/L} - i \left[ F_\mu^{R/L}, F_\nu^{R/L} \right].
\end{align}
The vector mesons are given in antisymmetric tensor representation and collected in the nonet matrix
\begin{align}
	V_{\mu\nu} = \begin{pmatrix} \rho^0_{\mu\nu} + \omega_{\mu\nu} & \sqrt{2} \rho^+_{\mu\nu} & \sqrt{2} K^+_{\mu\nu} \\
						 \sqrt{2} \rho^-_{\mu\nu} & -\rho^0_{\mu\nu} + \omega_{\mu\nu} & \sqrt{2}K^0_{\mu\nu} \\
						 \sqrt{2} K^-_{\mu\nu} & \sqrt{2} \bar{K}^0_{\mu\nu} & \sqrt{2} \phi_{\mu\nu} \end{pmatrix} \! .
\end{align}
Approximating the vector-meson masses by a common mass $\mv = 776\, \te{MeV}$, the vector-meson Lagrangian used in this article is given as \cite{Lutz:2008km, Terschlusen:2012xw}
\begin{align}
	&\Lvec = \Lag_{\te{free}} + \Lag_{\te{lin}}, \nn \adb \\
	&\Lag_{\te{free}} = -\frac{1}{4} \bra D^\mu V_{\mu\nu} \, D_{\rho}V^{\rho\nu} \ket + \frac{1}{8} \, \mv^2 \bra V_{\mu\nu} V^{\mu\nu} \ket , \nn \adb \\
	&\Lag_{\te{lin}} = \frac{1}{2} i f_V h_P \bra \Um_\mu V^{\mu\nu} \Um_\nu  \ket + \frac{1}{2} f_V \bra V^{\mu\nu} f^+_{\mu\nu} \ket \label{eq:Lvec}
\end{align}
with the still to be determined parameters $f_V$ and $h_P$ and the abbreviations
\begin{align}
	& D_\mu V_{\alpha\beta} := \partial_\mu V_{\alpha\beta} + \left[\Gamma_\mu, V_{\alpha\beta} \right], \nn \adb\\
	& \Gamma_\mu := \frac{1}{2} \left( \left[u^\dagger, \partial_\mu u \right] - i u^\dagger F_\mu^R u + i u F^L_\mu u^\dagger \right), \nn \adb \\
	& \Um_\mu := \frac{1}{2} u^\dagger D_\mu U u^\dagger = -\frac{1}{2} u \,D_\mu U^\dagger u, \nn \adb \\
	& f_{\mu\nu}^\pm := \frac{1}{2} \left(u F_{\mu\nu}^L u^\dagger \pm u^\dagger F_{\mu\nu}^R u \right), \nn \adb \\	
	& U = u^2\,.
\end{align}
With the particular choice of the kinetic terms given in \eqref{eq:Lvec}, the three vector-meson fields $V_{ik}$ for $i,k = 1, 2, 3$ are frozen, \textit{i.e.}, non-propagating fields \cite{Ecker:1988te}. For a different choice of the kinetic terms, other fields would be non-propagating. 

Note that the chiralised ``free'' Lagrangian $\Lag_{\te{free}}$ does contain interactions encoded in the chirally covariant derivative. The interactions between vector mesons and low-energy degrees of freedom are limited to $V$-$2P$ and $V$-$v$ couplings with vector mesons $V$, pseudoscalar mesons $P$ and an external vector source $v$, as already discussed in the introduction. These couplings describe the most prominent ways of interactions of vector mesons with pseudoscalar mesons. In particular, if one probes pions by the electromagnetic interaction, the pion form factor receives significant contributions from an intermediate $\rho$-meson, see, \textit{e.g.}, \cite{Leupold:2009, Terschlusen:2013iqa} and references therein. The next most significant terms, the $2V$-$P$ coupling \cite{Lutz:2008km, Terschlusen:2013iqa} and the mass splitting of the vector-meson masses \cite{Lutz:2008km, Terschlusen:2012xw} are not part of the present feasibility study. 
Note that the notation used within this article follows the one used in \cite{Terschlusen:2012xw} and differs from, \textit{e.g.}, the one used in \cite{Ecker:1988te}. In Tab.\ \ref{tab:CompNotation}, the corresponding notations are matched. 
\begin{table}[h]
\caption{Comparison between notations used in this article and in \cite{Ecker:1988te}. The latter are denoted by $\hat{\ }$ .}
\label{tab:CompNotation}
\begin{tabular}{l|l}
	notation in this article & notation in \cite{Ecker:1988te} \\ \hline 
	& \\[-0.5em]
	$\{ \Phi, \, V_{\mu\nu} \}$ & $\{ \hat{\Phi}, \, \hat{V}_{\mu\nu} \} = \frac{1}{\sqrt{2}} \{ \Phi, \, V_{\mu\nu} \}$ \\[0.5em]
	$ u^2= U = \exp( i \frac{\Phi}{F} ) $ & $ \hat{u}^2 = \hat{U} = \exp (-\sqrt{2} i \frac{\hat{\Phi}}{F}) = U^\dagger$ \\[0.5em]
	$D_\mu U = \partial_\mu  U - i F_\mu^R U$ & $\hat{D}_\mu \hat{U}  =  \partial_\mu U^\dagger - i F_\mu^R U^\dagger $ \\
	$ \phantom{D_\mu U = } +i U F_\mu^L$ &  $\phantom{\hat{D}_\mu \hat{U} = } +i  U^\dagger F_\mu^L$ \\[0.5em]
	$\Gamma_\mu = \frac{1}{2} ( [ u^\dagger, \partial_\mu u] - i u^\dagger F_\mu^R u $ & $\hat{\Gamma}_\mu = \frac{1}{2} ( [ u^\dagger, \partial_\mu u] - i u F_\mu^R u^\dagger$ \\
	$\phantom{\Gamma_\mu = \frac{1}{2} ( } - i u F_\mu^L u^\dagger )$ & $\phantom{ \hat{\Gamma}_\mu = \frac{1}{2} ( } - i u^\dagger F_\mu^L u)$ \\[0.5em]
	$\Um_\mu = - \Um_\mu^\dagger = \frac{1}{2} u^\dagger D_\mu U u^\dagger$ & $\hat{u}_\mu = \hat{u}_\mu^\dagger = i \hat{u}^\dagger \hat{D}_\mu \hat{U} \hat{u}^\dagger$ \\[0.5em]
	$f^{\pm}_{\mu\nu} = \frac{1}{2} (u F_{\mu\nu}^L u^\dagger \pm u^\dagger F_{\mu\nu}^R u)$ & $\hat{f}^{\pm}_{\mu\nu} = 2 (f^{\pm}_{\mu\nu})^\dagger$ \\[0.5em]
	$\chi_{\pm} := u^\dagger \chi u^\dagger \pm u \chi^\dagger u$ & $\hat{\chi}_{\pm} =  u \chi u \pm u^\dagger \chi^\dagger u^\dagger$ \\[0.5em]
	$\{h_P, \, f_V\}$ & $\{\hat{G}_V, \, \hat{F}_V\} = \{\frac{1}{4} f_V h_P , \, f_V\}$	
\end{tabular}
\end{table}

The generating functional to calculate one-loop contributions is given by
\begin{align*}
	&\te{e}^{iZ} = \te{e}^{i \int \df x \Lch^{\rm NLO}[\bar{U}]} \int \!\!\te{d} \mu [\{U,V\}] \te{e}^{i \int \df x \Lag_1}, \\
	&\Lag_1 := \Lch^{\rm LO} + \Lvec.
\end{align*}
The first integral describes tree-level diagrams up to $\order(Q^4)$ only so that it has to be evaluated at the classical solution $\Ub$ for pseudoscalar fields determined through the equation of motion (EOM) of the LO-$\chpt$ Lagrangian $\Lch^{\rm LO}$. Hereby, the vector-meson fields are treated as pure fluctuations, \textit{i.e.}, they do not contribute to the classical fields. The integral measure $\te{d} \mu$ denotes an integral over the pseudoscalar and vector fields $U(x)$ and $V_{\mu\nu}(x)$, respectively. 

To calculate the one-loop approximation, the field $U$ is expanded around its classical solution $\bar{U}$ as \cite{Gasser:1984gg}  
\begin{align}
	U &= \bar{u} \, \exp^{i\xi} \bar{u}, \ \ \bar{U} = \bar{u}^2, \label{eq:ExpU}
\end{align}
whereby $\xi$ is a traceless, hermitian matrix. Treating in addition the vector-meson fields $V$ as fluctuations yields a combined fluctuation vector $\hat{\xi} = (\xi, V)^t$. Therewith, $\Lag_1$ can be expanded in the neighbourhood of the classical solution $\Ub$. In that way, we define the matrix operator $D$ via
\begin{align}
	\int \df x  \Lag_1 [U] =:& \! \int \df x \Lag_1 [\Ub] - \frac{1}{2} \! \int \df x \, \df y  \hat{\xi}^t(x) D(x,y) \hat{\xi}(y) \nn \\
	&+ \order(\hat{\xi}^{\,3}). \label{eq:DefD}
\end{align}
The one-loop contribution can be expressed in terms of $D$ and, up to an irrelevant constant, is given by
\begin{align}
	Z_{\te{one loop}} &= \frac{1}{2} \, i \log (\det D)\,. \label{eq:Zoneloop}
\end{align}
Since the vector-meson fields are treated as pure fluctuations, the one-loop contribution depends only on the classical pseudoscalar fields and on external sources. Thus, all singularities therein have to have the structure of terms in the pure $\chpt$-Lagrangians and have to renormalise the low-energy constants therein such that the one-loop approximation for $Z$ is finite. In the present work we restrict ourselves to $\Lch^{\rm LO}$ and $\Lch^{\rm NLO}$.

\subsection{Determining the matrix $D$ for a general Lagrangian} \label{subsec:Dfree}

If one-loop contributions are calculated using Eq.\ \eqref{eq:Zoneloop}, the matrix $D$ defined according to Eq.\ \eqref{eq:DefD} is needed. It can be determined by expanding an action $Z {=} \int \df x  \mathscr{L}$ at the classical fields $\Ub$. Let $Z$ be a general action depending on fields $A_i$, $i {=} 1, \ldots, n $ for a given $n \in \mathbb{N}$. Then, the EOM of a field $A_j$ reads as
\begin{align*}
	0 = \left. \frac{\partial Z}{\partial A_j} \right|_{\{A_i \} = \{ \bar{A}_i \} }
\end{align*}
for all $j {=} 1, \ldots, n$. Hereby, $\{ \bar{A}_i \}$ denotes the classical fields. With the EOM, the action can be expanded and the matrix operator $D$ determined according to (integrations are implicit)
\begin{align}
	&Z[\{A_i \}] = Z[ \{\bar{A}_i\} ] - \frac{1}{2} \, \hat{\xi}^t D \hat{\xi} + \order(\xi^3)\,, \nn \adb \\
	&D_{ij}(x,y) = - \left. \frac{\partial Z}{\partial A_i(x) \partial A_j(y)} \right|_{\{A_k\} = \{ \bar{A}_k \} } . \label{eq:Def-D} 
\end{align}

\subsection{Expanding the one-loop contribution in powers of pseudoscalar and other external fields} \label{subsec:ExpInPhi}

In general, one is not only interested in how the low-energy constants are renormalised by the one-loop contribution including light vector mesons but also in their influence on observables like the pseudoscalar masses and decay constants (see further work by the same authors \cite{MassPaper}).
Thus, one might wonder whether one can determine the renormalisation of (some of) the low-energy constants by just calculating two-point functions, \textit{i.e.}, by expanding the one-loop functional up to second order in classical fields and/or external sources. However, there are several chiral structures up to $\order(Q^4)$ which contribute in the same way to (onshell) two-point functions. Therefore, only linear combinations of low-energy constants are related in this way to the infinities emerging from loops with vector mesons. To disentangle the impact of the loops on the various low-energy constants, one has to keep the complete chiral structure encoded in the field $U$ instead of expanding in powers of the fields. 

\subsection{Heat-kernel approach} \label{subsec:Heat-kernel}

Since the one-loop calculation including light vector mesons seems to be similar to the calculation with pseudoscalar mesons only, one could try to follow \cite{Gasser:1983yg, Gasser:1984gg} using a heat-kernel approach. In general, for using a heat-kernel approach a matrix $D$ according to the definition in Eq.\  \eqref{eq:DefD} is considered. This matrix has to fulfil the condition $D \rightarrow D_0 \sim \Box + (\te{mass})^2$ in the limit of no external fields. Here and in the following, the phrase ``limit of no external fields'' refers to the classical solution $\bar{u} \equiv 1$, the scalar source $s \equiv \mathcal{M}$ and all other external sources set to zero. Therewith, the matrix elements in $d$ dimensions can be expressed as \cite{Gasser:1983yg, Ball:1988xg}
\begin{align*}
	&\bra x \left| \te{e}^{- \lambda D} \right| y \ket =: \bra x \left| \te{e}^{- \lambda \Box} \right| y \ket H(x|\lambda |y) , \\ 
	&\bra x \left| \te{e}^{- \lambda \Box} \right| y \ket = i \left( 4 \pi \lambda \right)^{-d/2} \exp \left[ \frac{(x-y)^2}{4 \lambda} \right]
	\end{align*}
with a purely imaginary parameter $\lambda$. Then,
\begin{align}
	&\log (\det D) = - \operatorname{Tr} \int_0^{i \infty} \frac{\te{d}\lambda}{\lambda} \te{e}^{-\lambda D} \nn \adb \\
	& \ \ \ \ = -i \left( 4 \pi \lambda \right)^{-d/2} \int_0^{i \infty} \te{d}\lambda \, \lambda^{-(1+d/2)} \int \!\!\text{d}^d\! x  \bra H(x|\lambda |x) \ket .
\end{align}
After Taylor-expanding $H$ around $\lambda = 0 $, 
\begin{align}
	H(x|\lambda |y) = \sum_{n=0}^{\infty} \lambda^n H_n(x|y), 
\end{align}
the one-loop contribution is given by
\begin{align}
	\frac{i}{2} \log (\det D) 
	= - \frac{i}{2} \int \!\!\text{d}^d\! & x \left\{ \frac{1}{d}\, H_0(x|x) + \,\frac{1}{4 \pi (d{-}2)} \, H_1(x|x) \right. \nn \\
	&\left. \! \!+ \, \frac{1}{(4\pi)^2 (d{-}4)} \, H_2(x|x) \right\} + (\te{irrel.}) .
\end{align}
Therefore, only $H_2(x|x)$ has to be determined to identify the infinite contribution for the physical number of dimensions, $d {=} 4$. It can be determined using the differential equation for $H(x|\lambda |y)$ which is generated by taking the derivative of the matrix element with respect to $\lambda$,
\begin{align*}
	\frac{\partial}{\partial\lambda} \bra x \left| e^{-\lambda D} \right| y \ket  = - D_x \bra x \left| e^{-\lambda D} \right| y \ket ,
\end{align*}
and the initial condition $H(x|0|x) {=} H_0(x|x) {=} 1$. This differential equation yields recursive relations for the $H_n(x|y)$ which can be used to calculate $H_2(x|x)$. 

For loops including vector mesons in the antisymmetric tensor representation, the corresponding matrix $D$ does not have the required standard form $\Box + (\te{mass})^2$ in the limit of no external fields. However, a projection on the space of antisymmetric rank-2 tensors can be performed (cf.\ subsection \ref{subsec:Proj} for details on this projection) such that the vector fields are decomposed into a propagating mode and a non-propagating mode. \textit{I.e.}, in the limit of no external fields, the matrix $D$ is equal to $\Box + (\te{mass})^2$ if acting on the propagating mode and equal to $(\te{mass})^2$ only if acting on the non-propagating mode. It turns out that due to the non-standard form of the matrix $D$ acting on the non-propagating mode a heat-kernel approach is not applicable. This is discussed in greater detail in appendix \ref{app:heat-kernel} where the heat-kernel approach is applied to a toy Lagrangian with only one vector-meson flavour.

\subsection{Calculating one-loop contributions in powers of $D{-}D_0$} \label{subsec:SumN}

The heat-kernel method of \cite{Gasser:1984gg} is very elegant in providing a closed form $H_2$ for the divergences in four dimensions and in keeping chirally covariant structure throughout the calculation. In lack of this method, we have to resort to a more direct brute force approach. As we will see, this requires at some point a derivative expansion of a non-local expression with the aim of obtaining a local effective action. The {\em ordinary} derivatives that appear in this way must be fused in the end with the appropriate fields to obtain the pertinent chirally invariant structures that fit the Lagrangians of $\chpt$. This painful book-keeping procedure can, on the other hand, be seen as an important cross check of our calculations. It is a highly non-trivial check if several separately non-chiral terms fuse to chirally invariant structures.

To determine divergences in four dimensions, the calculation of one-loop contributions via an expansion in $\delta D {:=} D {-} D_0$ is discussed in this subsection. Hereby, $D_0$ denotes again the matrix $D$ in the limit of no external fields. Using $D {=} D_0 {+} \delta D$, the one-loop contribution can be rewritten as
\begin{align}
	&Z_{\te{one loop}} = \frac{1}{2} i \log \left[ \det (D_0 + \delta D) \right]  = \frac{1}{2} i \tr \left[ \log (D_0 + \delta D) \right] \nn \\
	& \ \ \ = \frac{1}{2} i \sum_{N=1}^{\infty} (-1)^{N+1} \, \frac{1}{N} \, \tr \left[\left( D_0^{-1} \, \delta D \right)^N \right] +  (\te{irrel.}) . \label{eq:SumN}
\end{align}
Thus, for an arbitrary $N \in \mathbb{N}$ one has to calculate
\begin{align}
	&\hspace{-1.5em} \tr \left[\left( D_0^{-1} \, \delta D \right)^N \right]  \nn \\
	 =& \prod_{i=1}^N \left\{ \int \df x_{2i-1} \ \df x_{2i} \int \frac{\df k_i}{(2\pi)^4} \exp \! \left[i k_i (x_{2i-1} - x_{2i}) \right] \right\}\nn \\
	 & \sd \bra \vphantom{x_{2j}} \right. \prod_{j=1}^{N} \left. D_0^{-1}(k_j) \delta D(x_{2j}, x_{2j+1}) \left. \vphantom{x_{2j}} \! \ket \right|_{x_{2N{+}1} := x_1}
\end{align}
Hereby, $D_0^{-1}$ denotes both the matrix in coordinate space and the corresponding one in momentum space. However, it is always clear from the context which one is used. 

As a first step, derivatives acting on $\delta$-functions which show up in $\delta D$ (see determination of $\Dvec$, $\Dch$ and $\Dmix$ in the following sections) have to be evaluated, 
\begin{align}
	\int \df x \, \te{e}^{ikx} A(x) \partial^x_\eta \delta(x-y) = - \te{e}^{iky} \left(i k_\eta + \partial_\eta \right) A(y) \,.
\end{align}
Next, the multidimensional space integral has to be localised, \textit{i.e.}, expanded around one space coordinate, \textit{e.g.}, around $x_1=:x$ with $x_i =: x {-} z_i$ for $i \neq 1$ and 
\begin{align*}
	A(x_i) = A(x - z_i) = \sum_{n=0}^{\infty} \frac{(-1)^n}{n!} z_i^{\mu_1} \! \dotsm z_i^{\mu_n} \, \partial_{\mu_1} \! \dotsm \partial_{\mu_n} A(x) \,.
\end{align*}
Since $\partial \in \order(Q)$, this Taylor expansion can be approximated by a finite series if calculating $\tr(\log D)$ to a given order in $Q$. Note that at this point the ordinary derivatives appear which have to be fused with appropriate fields in the end to obtain chirally invariant structures.

If the integrand is proportional to the exponential $e^{ik z_i}$ for a given momentum $k$ and space point $z_i$ after the transformation above, it will not be proportional to the exponential $e^{ik z_j}$ for the same momentum $k$ and another space point $z_j \neq z_i$. Furthermore, no exponential function in the integrand depends on the expansion point $x$. Thus, after the transformation
\begin{align*}
	\int \! \frac{\df k}{(2\pi)^4} \, z_i^\mu \te{e}^{ik z_i} B(k) = i \! \int \! \frac{\df k}{(2\pi)^4} \, \te{e}^{ikz_i} \partial_\mu B(k)
\end{align*}
all integrals $\int \df z_i$ can be performed yielding $\delta$-functions for the momentum variables. The evaluation of those $\delta$-functions reduces $\bra (D_0^{-1} \delta D)^N \ket$ to an integral over both one space and one momentum variable only.

To identify the infinite part of the momentum integral, dimensional regularisation is used, \textit{i.e.}, the integral in momentum space is calculated in $(4{+}2\varepsilon)$ instead of four dimensions. Its integrand can be further simplified containing one propagator with a common mass instead of several propagators with separate masses using Feynman parameters \cite{Bailin-Love}, 
\begin{align}
	&\hspace{-1.5em} \left( a_1^{\alpha_1} \dotsm a_n^{\alpha_n} \right)^{-1} \nn \\
	=& \frac{\GammaF(\sum \alpha_i)}{\prod \GammaF(\alpha_i)} \int_0^1 \te{d} u_1 \int_0^{u_1} \te{d} u_2 \dotsm \int_0^{u_{n-2}} \te{d} u_{n-1} \nn \\
	& \sd \left\{ \frac{(1-u_1)^{\alpha_1 -1} (u_1 - u_2)^{\alpha_2 -1} \dotsm u_{n-1}^{\alpha_n -1}}{\left[a_1 (1-u_1) + a_2 (u_1 - u_2) + \dots + a_n u_{n-1} \right]^{\sum \alpha_i}} \right\} .
\end{align}
In $(4 + 2\varepsilon)$ dimensions, a momentum integral with one propagator is given by \cite{Pascual:1984zb}
\begin{align}
	&\frac{1}{\mu^{2\eps}} \int \frac{\!\! \text{d}^{4{+}2\eps} k}{(2\pi)^{4{+}2\eps}} \frac{[k^2]^\alpha}{[k^2-m^2 + i\eta]^\beta} \nn \\
	& = \frac{i}{16\pi^2} (-m^2)^{\alpha{-}\beta{+}2} \left(\frac{m^2}{4\pi\mu^2} \right)^{\!\!\eps} \frac{\GammaF(2{+}\alpha{+}\eps) \GammaF(-\alpha{+}\beta{-}2 {-}\eps)}{\GammaF(\beta) \GammaF(2{+}\eps)} \nn \adb \\
	& =  \frac{i \, f(m^2,\alpha{-}\beta)}{16\pi^2} \left( \frac{1}{\varepsilon} + \GammaF^{\prime}\!(1)-1 - \log(4\pi) \right)  + (\te{finite}) \nn \adb \\
	& =: i f(m^2,\alpha{-}\beta) \cdot \bar{\lambda} + (\te{finite})  \label{eq:FormularMomInt}
\end{align}
for a small $\eps \in \mathbb{R}$, $\alpha \in \mathbb{N}_0$ and  $\beta \in \mathbb{N}$. The finite part consists of terms of $\order(1)$ and terms of $\order(\varepsilon)$ which vanish for $\varepsilon {\rightarrow} 0$. The function $f: \mathbb{R}^2 \rightarrow \mathbb{R}$ depends only on the mass $m$ and the combination $(\alpha{-}\beta)$ but not on $\eps$. Indeed, $f(\bullet, \alpha{-} \beta) \equiv 0$ for $(\beta{-}\alpha) \ge 3$, \textit{i.e.}, the integral is finite. The renormalisation scale $\mu$ is introduced by dimensional regularisation. Note that all physical observables have to be independent of the scale $\mu$.

Furthermore, if an integrand of the form given in the integral above is multiplied with $k^{\mu_1} \dotsm k^{\mu_n}$, the integral will be zero for all odd $n \in \mathbb{N}$. Otherwise, the multiplicand can be substituted by \cite{Pascual:1984zb}
\begin{align}
	&k^{\mu_1} k^{\mu_2} \mapsto \frac{k^2}{4 {+} 2 \eps} \, g^{\mu_1 \mu_2}, \nn \\
	&k^{\mu_1} \dotsm k^{\mu_4} \mapsto \frac{k^4}{(4 {+} 2 \eps) (6 {+} 2 \eps) } \left(g^{\mu_1 \mu_2} g^{\mu_3 \mu_4} + g^{\mu_1 \mu_3} g^{\mu_2 \mu_4} \right. \nn \\[-1.5em]
	& \phantom{k^{\mu_1} \dotsm k^{\mu_4} \mapsto \frac{k^4}{(4 {+} 2 \eps) (6 {+} 2 \eps) } } \left. \, +\,  g^{\mu_1 \mu_4} g^{\mu_2 \mu_3} \right) \label{eq:Repl-kmu}
\end{align}
and accordingly for $n > 4$, $n$ even.

In section \ref{app:Ex-int} in the appendix, an integral is calculated as an example for the procedure described in this subsection.

\section{One-loop contributions including vector mesons up to $\order(Q^4)$} \label{sec:Octet}

In this section, the one-loop contribution including vector mesons are calculated up to $\order(Q^4)$ (subsections \ref{subsec:vec}-\ref{subsec:mix}). The calculation method is based on the techniques discussed in subsection \ref{subsec:SumN}. Furthermore, the results are used to renormalise the low-energy constants of the LO- and NLO-$\chpt$ Lagrangians \eqref{eq:ChPTLagr} (see subsection \ref{subsec:Renorm}).

For fluctuations both in the pseudoscalar and in the vector-meson fields as considered in this article, the matrix $D$ can be written as a block matrix such that 
\begin{align}
	&\hat{\xi}^t D \hat{\xi} = (\xi^t, V^t) \begin{pmatrix} \Dvec & \Dmix \\ \Dmix^{t} & \Dch \end{pmatrix} \begin{pmatrix} \xi \\ V \end{pmatrix} \nn \\
	& \ = V^t \Dvec V + V^t \Dmix \xi + \xi^t \Dmix^{t} V + \xi^t \Dch \, \xi\,.
\end{align}
Using this block structure, equation \eqref{eq:SumN} which expresses the one-loop contribution as a sum over $D_0$ and $\delta D$ with $D{=} D_0 + \delta D$ can be split up into parts containing or not containing $\Dmix$, respectively,
\begin{align}
	\tr(\log D) &= \phantom{=:} \hspace{-1em} \tr(\log\Dvec) + \tr(\log\Dch) + \begin{pmatrix} \te{parts with} \\ \Dmix \end{pmatrix} \nn \\
	 &=: \phantom{=} \hspace{-1em} \tr(\log\Dvec) + \tr(\log\Dch) + \tr(\log D)_{\te{mix}}.
\end{align}
The different parts of this sum are calculated separately in subsections \ref{subsec:vec} - \ref{subsec:mix}. First, the one-loop contribution from $\Dvec$ is calculated, then the additional contribution from $\Dch$ and at last the contributions containing $\Dmix$. Thereby, all calculations are performed up to $\order(Q^4)$. Furthermore, the projection on the space of antisymmetric rank-2 tensors necessary in order to determine $\Dvec^0$ is discussed in the following subsection \ref{subsec:Proj}.

\subsection{Projection on the space of antisymmetric rank-2 tensors} \label{subsec:Proj}
As discussed in subsection \ref{subsec:SumN}, the limit $D_0$ for no external fields has to be determined in order to calculate the one-loop contribution. If the matrix $D$ is written as a block matrix, this limit has to be determined for all block-matrix parts separately. As already calculated in \cite{Gasser:1984gg}, $\Dch^0 \sim \Box + (\te{mass})^2$. Furthermore, $\Dmix^0 = 0$. For determining $\Dvec^0$ consider the free Lagrangian $\Lfree$ given in Eq. \eqref{eq:Lvec} evaluated at the classical solution of the pseudoscalar fields, $\Ub = \bar{u}^2$,  
\begin{align}
	&\int\df x \Lag_{\te{free}}|_{U = \bar{U}} \nn \\
	&= - \frac{1}{4} \int \df x \ \df y \, V_{\mu\nu}^a(x)\, \Dvec(x,y)^{\mu\nu\alpha\beta}_{ab}\,V^b_{\alpha\beta}(y)\,.
\end{align}
Since $\Dvec$ is generated by the parts in the Lagrangian containing two vector meson fields, it is generated by $\Lag_{\te{free}}$ only. Hereby, the matrix $\Dvec$ is twice the definition in Eq. \eqref{eq:Def-D} in order to simplifying further calculations. This only adds a constant to $\tr(\log\Dvec)$ and, hence, does not change the final result.  In the following, the matrices $\Dch$ and $\Dmix$ are determined in the same way. 

In the limit of all external fields set to zero,
\begin{align}
	&\Dvec(x,y)^{\mu\nu\alpha\beta} \rightarrow \Dvec^0(x,y)^{\mu\nu\alpha\beta} \nn\\ 
	&= - \! \left( 2 P_1^{\mu\nu\tau\rho} P_1^{\alpha\beta\eta\sigma}g_{\rho\sigma}  \partial^x_{\tau} \partial^x_\eta + \mv^2 P_1^{\mu\nu\alpha\beta} \right) \delta(x-y)
\end{align}
including the unit element of the vector space of all antisymmetric rank-2 tensors,
\begin{align}
	P_1^{\mu\nu\alpha\beta} := \frac{1}{2} \left( g^{\mu\alpha} g^{\nu\beta} - g^{\mu\beta} g^{\nu\alpha} \right). 
\end{align}
Since the vector meson fields $V_{\mu\nu}$ are antisymmetric tensor fields, $\Dvec$ only acts on the space of antisymmetric rank-2 tensors. Hence,  $\Dvec$ can be reduced explicitly to a matrix over the vector space of antisymmetric tensor fields without changing the result of $\tr(\log \Dvec)$. Therefore, the antisymmetric projection operators in momentum space,
\begin{align}
	& \widetilde{P}_V^{\mu\nu\alpha\beta}(k) := \frac{1}{2k^2} \left(g^{\mu\alpha} k^\nu k^\beta - g^{\mu\beta} k^\nu k^\alpha - g^{\nu\alpha} k^\mu k^\beta \right. \nn \\[-0.5em]
	& \phantom{\widetilde{P}_V^{\mu\nu\alpha\beta}(k) := 2k^2 \left( \right.} \left. +\, g^{\nu\beta} k^\mu k^\alpha \right) \, , \nn \\
	& \widetilde{P}_A := \widetilde{P}_1-\widetilde{P}_V,\, \widetilde{P}_A^2 = \widetilde{P}_A, \, \widetilde{P}_V^2 = \widetilde{P}_V, \, \widetilde{P}_A \bot \widetilde{P}_V \, \label{eq:Def:proj-operators}
\end{align}
are introduced. Reduced to the antisymmetric space, the matrix $\Dvec \mapsto P \Dvec P$ with $P = \te{diag} \left\{P_A, P_V \right\}$ and the projection operators $P_{A/V}$ in coordinate space. Then, 
\begin{align}
	\tr(\log \Dvec) &= \tr [\log \left( P \Dvec P \right)] \nn \\
	& \hspace{-2.2em} = \sum_{N=1}^{\infty} (-1)^{N+1} \frac{1}{N} \tr \left[ \left( P [\Dvec^0]^{-1} P \sd P \delta \Dvec P  \right)^N \right] \nn \\
	& \hspace{-2.2em} 	= \sum_{N=1}^{\infty} (-1)^{N+1} \frac{1}{N} \tr \left[ \left( P [\Dvec^0]^{-1} \sd \delta \Dvec  \right)^N \right] 
\end{align}
since $[\Dvec^0]^{-1} P {=} P [\Dvec^0]^{-1}$ and  $P^2 {=} P$. Hence, the only matrix which actually has to be reduced to the antisymmetric space is $\Dvec^0$. Then, the inverted matrix of the reduced matrix $P \Dvec^0$ is equal to
\begin{align}
	P [\Dvec^0]^{-1}= -\left( \Box + \mv^2 \right)^{-1} P_V - \mv^{-2} P_A.
\end{align}
As can be seen here, the operator $P_V$ projects on the propagating vector-meson mode while the operator $P_A$ projects on the non-propagating mode.

The mixed matrix operator $\Dmix$ acts both on antisymmetric vector fields $V_{\mu\nu}$  and pseudoscalar fields $U$. The part acting on vector fields is multiplied with $\Dvec^0$ in all further calculations. Again, $\Dvec^0$ and therewith also the corresponding part of $\Dmix$ have to be reduced explicitly to matrices over the vector space of antisymmetric tensor fields to achieve the desired form of $\Dvec^0$. Hereby, the reduced matrix $\Dmix$ is equal to $P \Dmix$ and $\Dmix^t P$, respectively. Then,
\begin{align*}
	&P [\Dvec^0]^{-1}] P \sd P \Dmix \sd  \Dch^{-1} = P [\Dvec^0]^{-1} \sd  \Dmix \sd \Dch^{-1}, \\
	& \Dch^{-1} \sd \Dmix^{t} P \sd P [\Dvec^0]^{-1} P =  \Dch^{-1} \sd \Dmix^{t} \sd P [\Dvec^0]^{-1}.
\end{align*}
Hence, also for terms including $\Dmix$ in $\tr(\log D)$ it is sufficient to only reduce $[\Dvec^0]^{-1}$ to $P [\Dvec^0]^{-1}$.

\subsection{Result for $\tr(\log \Dvec)$} \label{subsec:vec}
As discussed before, $\Dvec$ is generated by the parts in the Lagrangian containing two vector meson fields with all pseudoscalar fields evaluated at their classical solution $\bar{U} = \bar{u}^2$. It can be decomposed as\footnote{Recall from the previous subsection that $P[\Dvec^0]^{-1}$ is needed to calculate $\tr\left( \log \Dvec \right)$ instead of $[\Dvec^0]^{-1}$ only.}
\begin{align}
	& \Dvec =: P \Dvec^0 + \Delta, \adb \nn \\
	& \Delta(x,y)^{\mu\nu\alpha\beta}_{ab} =: [ F(x,y)^{\mu\nu\alpha\beta}_{ab} + H(x,y)^{\mu\nu\alpha\beta, \eta}_{ab}\partial^x_\eta ] \delta(x-y)
\end{align}
with $\Delta$ containing both a local term, $F$,  and a term with an additional derivative, $H \partial$. The matrices $F$ and $H$ are both antisymmetric in the Lorentz indices $(\mu,\nu,\alpha,\beta)$,
\begin{align}
	& F(x,y)^{\mu\nu\alpha\beta}_{ab} := P_1^{\mu\nu\tau\rho} P_1^{\alpha\beta\eta\bar{\rho}} E(x,y)^{ab}_{\tau\eta} \, g_{\rho \bar{\rho}}, \nn \adb \\
	& H(x,y)^{\mu\nu\alpha\beta, \eta}_{ab} := P_1^{\mu\nu\tau\rho} P_1^{\alpha\beta\eta\bar{\rho}} G(x)^{ab}_\tau g_{\rho \bar{\rho}} + \begin{pmatrix} \mu\nu \leftrightarrow \alpha\beta, \\ x \leftrightarrow y \end{pmatrix}. \label{eq:DefFH}
\end{align} 
Finally, the matrices $E$ and $G$ contain the flavour information of $\Delta$ and the building blocks of the Lagrangian directly, 
\begin{align}
	& E(x,y)^{ab}_{\tau\eta} := \bra [\Gamma_\tau(x), \lambda^a] [\Gamma_\eta(y), \lambda^b ] \ket = E(y,x)^{ba}_{\eta\tau}, \nn \adb \\
	& G(x)^{ab}_\tau := \bra[ \lambda^a, \lambda^b ] \Gamma_\tau(x) \ket = - G(x)^{ba}_\tau \label{eq:DefEG}
\end{align}
with $\lambda^0 {:=} \sqrt{2/3} \cdot \mathbb{1}$ and the Gell-Mann matrices $\lambda^1, \dotsc, \lambda^8$. 

As $H \in \order(Q)$, the one-loop contribution from $\Dvec$ up to $\order(Q^4)$ is given by the finite sum
\begin{align*}
	&\tr(\log D_{\te{vec}}) = \sum_{N=1}^{4}  \, \frac{(-1)^{N+1}}{N} \, \tr [\left(( P D_0^{-1} \cdot \Delta  \right)^N ] + \order(Q^6) \\
	&\phantom{\tr(\log D_{\te{vec}}) } \ =: i \bar{\lambda} \int\df x \, Q_4^{\te{vec}} + \te{(finite)} + \order(Q^6)
\end{align*}	
with $\bar{\lambda}$ as defined in Eq.\ \eqref{eq:FormularMomInt} and 	
\begin{align}
	&Q_4^{\te{vec}} = -\frac{3}{2} \, \mv^2 \bra G^2  + 2 E^\tau_\tau \ket - \frac{1}{128} \left\{ 16 \bra (\partial G)^2 + G \sd \Box G\ket  \right. \nn \\
	&\phantom{Q_4^{\te{vec}} = \,}  - 32 \bra \partial^\tau G^\eta G_\tau G_\eta \ket - 10 \bra  (G^2)^2 \ket + 13 \bra (G_\tau G_\eta)^2 \ket \nn \\
	&\phantom{Q_4^{\te{vec}} = \,} + 12\bra E^\tau_\tau G^2  + E^{\tau\eta} [G_\tau , G_\eta ] \ket -128 \bra \partial_\tau G_\eta E^{\tau\eta} \ket  \nn \\
	&\phantom{Q_4^{\te{vec}} = \,} \left. + \, 12 \bra (E^\tau_\tau)^2 \ket + 4 \bra E_{\tau\eta} (E^{\tau\eta}-E^{\eta\tau}) \ket \right\}	. 
\end{align}
Be aware that there are two types of traces involved in $Q_4^{\te{vec}}$. Both $E$ and $G$ are $9 \times 9$ matrices in flavour space. However, according to the definition in Eq.\ \eqref{eq:DefEG} each component of $E$ and $G$, respectively, is given by a trace over $3 \times 3$ matrices. If the traces in flavour space in $Q_4^{\te{vec}}$ are rewritten component-by-component, the involved traces of $3 \times 3$ matrices can be calculated using \cite{Gasser:1984gg}
\begin{align}
	&\sum_{a=0}^8 \bra \lambda^a A \lambda^a B \ket = 2 \! \bra A \ket \! \bra B \ket, \ \sum_{a=0}^8 \bra \lambda^a A \ket \! \bra \lambda^a B \ket = 2 \! \bra A B \ket. \nn \\[-1em] \label{eq:GellMannSum}
\end{align}
Therewith, it is easy to see that the $\order(Q^2)$ contribution of $Q_4^{\te{vec}}$ is vanishing since $\bra G^2 \ket {=} {-}2 \bra E^\tau_\tau \ket$. With the field strength tensor $\Gamma_{\mu\nu} {:=} \partial_\mu \Gamma_\nu {-} \partial_\nu \Gamma_\mu {+} [\Gamma_\mu, \Gamma_\nu]$, the full  result for $Q_4^{\te{vec}}$ can then be expressed as
\begin{align}
	Q_4^{\te{vec}} =&\ 0 + \order(Q^4) =  -\frac{3}{2} \bra \Gamma_{\mu\nu} \Gamma^{\mu\nu} \ket \nn \\
	=&  -\frac{3}{32} \bra D_\mu \Ub^\dagger \, D^\mu \Ub \ket ^2 	- \frac{3}{16} \bra D_\mu \Ub^\dagger \, D_\nu \Ub \ket^2 \nn \\
	& + \frac{9}{16} \bra (D_\mu \Ub^\dagger \, D^\mu \Ub)^2 \ket + \frac{3}{4} \bra F^R_{\mu\nu} \bar{U} F_L^{\mu\nu} \bar{U}^\dagger \ket \nn \\
	& + \frac{3}{4}i \bra F_R^{\mu\nu} D_\mu \Ub \, D_\nu \Ub^\dagger + F_L^{\mu\nu} D_\mu \Ub^\dagger \, D_\nu \Ub \ket \nn \\
	& + \frac{3}{8} \bra F_R^{\mu\nu} F^R_{\mu\nu} + F_L^{\mu\nu} F^L_{\mu\nu} \ket. \label{eq:resDvec}
\end{align}
Hereby, the relation \cite{Gasser:1984gg}
\begin{align}
	\bra (D_\mu \Ub^\dagger \,  D_\nu \Ub)^2 \ket  =& \,  \frac{1}{2} \bra D_\mu \Ub^\dagger \, D^\mu \Ub \ket^2 + \bra D_\mu \Ub^\dagger \, D_\nu \Ub \ket^2 \nn \\
	& - 2 \bra (D_\mu \Ub^\dagger \, D^\mu \Ub)^2 \ket \label{eq:RefDmu}
\end{align}
was used. We also took from \cite{Gasser:1984gg} the matching of $\bra \Gamma_{\mu\nu} \Gamma^{\mu\nu} \ket$ to the form in which the NLO Lagrangian $\Lch^{\rm NLO}$ is displayed there. The contributions from $\Dvec$ renormalise the low-energy constants $L_1$, $L_2$, $L_3$, $L_9$, $L_{10}$ and $H_1$ of the NLO-$\chpt$ Lagrangian (see subsection \ref{subsec:Renorm}).

\subsection{Result for $\tr(\log \Dch)$} \label{subsec:ChPT}

$\Dch$ is generated by terms in the Lagrangian proportional to $\xi^2$. As will be shown in the following, all three parts of the Lagrangian, $\Lag_\chpt$, $\Lag_{\te{free}}$ and $\Lag_{\te{lin}}$, can contribute to $\Dch$. The contribution generated only by $\Lag_\chpt$ was already calculated in \cite{Gasser:1984gg}. We have used these results to successfully check our calculation method. However, this calculation is not presented in this article. 

The Lagrangians $\Lfree$ and $\Llin$ do not directly depend on the matrix $U$ describing the pseudoscalar fields but on the matrix $u = \sqrt{U}$. However, the expansion rule \eqref{eq:ExpU} for expanding $U$ at its classical solution $\Ub$ cannot be reformulated easily as an expansion rule for $u$. Therefore, the vector fields $V$ are rewritten such that $\Lfree$ and $\Llin$ depend on $U$ directly by introducing the fields $\widetilde{V} {:=} u V u^\dagger$. In terms of $\widetilde{V}$, the free vector Lagrangian reads as
\begin{align}
	& \Lfree = \frac{1}{4} \bra \widetilde{D}^\mu \widetilde{V}_{\mu\nu} \, \widetilde{D}_{\rho}\widetilde{V}^{\rho\nu} \ket + \frac{1}{8} \, \mv^2 \bra \widetilde{V}_{\mu\nu} \widetilde{V}^{\mu\nu} \ket , \nn \\
	& \widetilde{D}^\mu \widetilde{V}_{\mu\nu} := \partial_\mu \widetilde{V}_{\mu\nu} + [ \widetilde{\Gamma}^\mu, \widetilde{V}_{\mu\nu} ], \nn \\
	& \widetilde{\Gamma}^\mu := -\frac{1}{2} \partial^\mu U \, U^\dagger - \frac{1}{2} i \left( r^\mu + U l^\mu U^\dagger \right) = u \Gamma^\mu u^\dagger - \partial^\mu u \, u^\dagger
\end{align}
and the linear one as
\begin{align}
	& \Llin = \frac{1}{8} i f_V h_P \bra \widetilde{V}^{\mu\nu} D_\mu U \,  D_\nu U^\dagger \ket + \frac{1}{2} f_V \bra \widetilde V^{\mu\nu} \widetilde{f}^+_{\mu\nu} \ket, \nn \\
	& \widetilde{f}^+_{\mu\nu} := \frac{1}{2} \left( U F^L_{\mu\nu} U^\dagger + F^R_{\mu\nu} \right) = u f_{\mu\nu}^+ u^\dagger \,.
\end{align}
The fluctuation vector $\hat{\xi}$ is replaced by the transformed fluctuation vector $\{\widetilde{V}, \xi\}$ and can be treated in the same way as the original one in all calculations. Thereby, the differential transforms as
\begin{align}
	\prod_{i,j} \te{d} V_{ij} = \left[ \det (u) \det (u^\dagger) \right]^{-N_f} \prod_{i,j} \te{d} \widetilde{V}_{ij} = \prod_{i,j} \te{d} \widetilde{V}_{ij} \nn
\end{align}
with the number of flavours $N_f = 3$. Using this transformation, one can rewrite
\begin{align}
	\tr(\log D) = \tr (\log \widetilde{D}) .
\end{align}
In particular, the result for $\tr(\log \Dvec)$ \eqref{eq:resDvec} calculated in the previous subsection does not change for $V \mapsto \widetilde{V}$. 

The vector meson fields $\widetilde{V}$ have to be evaluated at their classical solution $\Vt_{\te{cl}}$ to get the terms in the Lagrangian quadratic in the fluctuations $\xi$ of the pseudoscalar field. $\Vt_{\te{cl}}$ is the solution of the EOM generated by the Lagrangians with vector mesons, $\Lfree$ and $\Llin$, 
\begin{align}
	0 =\,& - \big( \widetilde{D}_{\te{vec}}^0 + \widetilde{F} + \widetilde{H}^\eta \partial_\eta - \frac{1}{2} \partial_\eta \widetilde{H}^\eta \big)_{\mu\nu\alpha\beta}^{ab} \, \big[\Vt_{\te{cl}} \big]^{\alpha\beta}_b \nn \\
	&+ \frac{1}{8} i f_V h_P \bra \lambda^a D_\mu \Ub \,  D_\nu \Ub^\dagger \ket + \frac{1}{2} f_V \bra \lambda^a \widetilde{f}^+_{\mu\nu} \ket
\end{align}
evaluated at the classical solution $\Ub$ for the pseudoscalar fields. Here, $\widetilde{F}$ and $\widetilde{H}$ are defined as in Eq. \eqref{eq:DefFH} but with $\widetilde{\Gamma}$ instead of $\Gamma$. The classical field can be determined order by order as a solution of the EOM in the corresponding order, \textit{i.e.}, $\Vt_{\te{cl}} = \Vt_{\te{cl}}^{(0)} + \Vt_{\te{cl}}^{(2)} + \order(Q^4)$. At $\order(Q^0)$, the classical field is equal to zero since the EOM at $\order(Q^0)$ is given by
\begin{align}
	0 = \frac{1}{4} \, \mv^2 \Vt_{\te{cl}}^{(0)} .
\end{align}
Thus, the classical field is of $\order(Q^2)$ and its LO contribution is given as\footnote{In the following subsection, $\Vt$ has to be split according to $\Vt=P_V\Vt+P_A\Vt$. For $\order(Q^4)$ and higher, the classical solution has to be calculated separately for $P_V \Vt$ and $P_A \Vt$ since $\widetilde{D}_{\te{vec}}^0$ acts for higher orders differently on both parts.}
\begin{align}
[\Vt^{(2)}_{\te{cl}}]^a_{\mu\nu} = - \frac{f_V}{32\, \mv^2} \Big(i h_P \bra \lambda^a D_\mu \Ub \, D_\nu \Ub^\dagger \ket  + 4 \bra \lambda^a \widetilde{f}^+_{\mu\nu} \ket \Big) \label{eq:Def-Vcl}
\end{align}
for $\widetilde{f}^+_{\mu\nu}$ evaluated at the classical field $\Ub$. Note that in the present work we are interested in the effective Lagrangian where vector mesons are completely integrated out. Thus, the solution of the EOM for the vector-meson fields is the one where the homogeneous solution is put to zero and the inhomogeneous solution is purely caused by the source terms encoded in $\Lag_{\te{lin}}$ \eqref{eq:Lvec}.

If the vector Lagrangians are evaluated at the classical solution $\Vt_{\te{cl}}$, the Lagrangian $\Lfree$ will be quadratic in $\Vt_{\te{cl}}$ while in the Lagrangian $\Llin$ $\Vt_{\te{cl}}$ will always appear together with a block of $\order(Q^2)$. Therefore, $\Lfree {+} \Llin$ evaluated at the classical solution $\Vt_{\te{cl}}$ is a chiral invariant Lagrangian of $\order(Q^4)$ in the pseudoscalar fields since $\Vt_{\te{cl}} \in \order(Q^2)$, \textit{i.e.}, it has the same form as the $\chpt$-Lagrangian of $\order(Q^4)$, $\Lch^{\rm NLO}$. Hence, it cannot contribute to the one-loop contributions at $\order(Q^4)$ because the $\chpt$-Lagrangian $\Lch^{\rm NLO}$ does not contribute, either. Thus, $\tr(\log \Dch)$  up to $\order(Q^4)$ is determined by the pure $\chi$PT Lagrangian $\Lch^{\rm LO}$ only. This contribution was already calculated in \cite{Gasser:1984gg} renormalising all low-energy constants in $\Lch^{\rm NLO}$ except $L_3$ and $L_7$.

\subsection{Result for $\tr(\log D)$ containing $\Dmix$} \label{subsec:mix}
$\Dmix$ is determined from terms in the Lagrangian containing both one vector-meson field $\Vt$ as a fluctuation and one fluctuation $\xi$ in the pseudoscalar fields, \textit{i.e.}, from both the Lagrangian $\Lfree$ and $\Llin$. Thereby, one vector-meson field $\Vt$ in $\Lfree$ is taken as a fluctuation and the other one is replaced by the classical field $\Vt_{\te{cl}}$ given in \eqref{eq:Def-Vcl}. Since there are no terms involving $\Dmix$ in the first term $(N {=}1)$ of the series \eqref{eq:SumN}, $\Dmix$ is only needed up to $\order(Q^3)$ to calculate one-loop contributions up to $\order(Q^4)$. Additionally, $\Dmix \rightarrow 0$ in the limit of no external fields. $\Dmix$ is given by\footnote{Recall that $\Dmix$ is twice the definition in Eq. \eqref{eq:Def-D} (cf. subsection \ref{subsec:Proj}).}
\begin{align}
	&\Dmix (x,y) = [ L(x) + K_\eta(x)\partial_x^\eta + J_{\tau\eta}(x) \partial_x^\tau \partial_x^\eta ] \delta(x-y), \nn \\
	& L^{\mu\nu} = \frac{1}{4} f_V h_P \left(\gamma - 2 \delta \right)^{\mu\nu} + \frac{1}{2} i f_V \zeta^{\mu\nu} , \nn \adb \\
	& K^{\mu\nu}_{\eta} = \frac{1}{2} f_V h_P P_1^{\mu\nu\alpha\beta} g_{\alpha \eta } \vartheta_\beta + \frac{f_V}{2 \mv^{2}} \left\{- [ h_p \iota + 2i \kappa ]^\tau_{\alpha\beta} P_{\tau\eta}^{\mu\nu\alpha\beta} \right. \nn \\
	& \phantom{K^{\mu\nu}_{\eta} = } \left. + [ h_P \left( \varphi + d\gamma \right) + 2i \left( \psi + d\omega \right) ]^\tau_{\alpha\beta} \left(P_{\tau\eta}^{\mu\nu\alpha\beta}+P_{\eta\tau}^{\mu\nu\alpha\beta} \right) \right\}\! , \nn \adb \\
	& J^{\mu\nu}_{\tau\eta} = \frac{1}{2} f_V \mv^{-2} \left\{ h_p \gamma + 2i \omega \right\}_{\alpha\beta} P_{\tau\eta}^{\mu\nu\alpha\beta}
\end{align}
with the abbreviations
\begin{align}
	& P_{\tau\eta}^{\mu\nu\alpha\beta} := P_1^{\mu\nu\rho\sigma} P_1^{\alpha\beta\bar{\rho}\bar{\sigma}} g_{\tau \rho} g_{\eta \bar{\rho}} g_{\sigma \bar{\sigma}}, \nn \adb \\
	& \gamma_{\alpha\beta}^{ab} := \bra [\ub^\dagger \lambda^a \ub, \lambda^b ] [\Umb_\alpha, \Umb_\beta ] \ket , \, d\gamma^{\tau, ab}_{\alpha\beta} := \gamma_{\alpha\beta}^{ab}|_{[\Umb,\Umb] \mapsto \partial^\tau [\Umb,\Umb]} \nn \adb \\
	& \delta_{\alpha\beta}^{ab} := P_{1, \alpha\beta\rho\sigma} \bra [ \ub^\dagger \lambda^a \ub, \Umb^\rho ] [ \lambda^b, \Gamma^\sigma] \ket , \nn \adb \\
	& \zeta_{\alpha\beta}^{ab} := \bra [\ub^\dagger \lambda^a \ub, \lambda^b ] \ub F^L_{\alpha\beta} \ub^\dagger \ket , \nn \adb \\
	& \vartheta^{ab}_\beta := \bra [ \ub^\dagger \lambda^a \ub, \lambda^b ] \Umb_\beta \ket , \nn \adb \\
	& \iota^{\tau,ab}_{\alpha\beta} := \bra [ \ub^\dagger \lambda^a \ub, [ \Umb_\alpha, \Umb_\beta ] ] [ \lambda^b, \Umb^\tau ] \ket , \nn \adb \\
	& \kappa^{\tau,ab}_{\alpha\beta} := \bra [ \ub^\dagger \lambda^a \ub, f^+_{\alpha\beta} ] [ \lambda^b, \Umb^\tau ] \ket , \nn \adb \\
	& \varphi^{\tau,ab}_{\alpha\beta} := \bra [ \ub^\dagger \lambda^a \ub , \Gamma^\tau] [\lambda^b ,[\Umb_{\alpha}, \Umb_\beta ] ] \ket , \nn \adb \\
	& \psi^{\tau,ab}_{\alpha\beta} := \bra [ \ub^\dagger \lambda^a \ub , \Gamma^\tau] [\lambda^b , f^+_{\alpha\beta} ] \ket , \nn \adb \\
	& \omega_{\alpha\beta}^{ab} := \bra [ \ub^\dagger \lambda^a \ub, \lambda^b ] f_{\alpha\beta}^+ \ket , \, d\omega^{\tau,ab}_{\alpha\beta} := \omega_{\alpha\beta}^{ab}|_{f^+ \mapsto \partial^\tau f^+} , \nn \adb \\
	&\Umb_\alpha := \frac{1}{2} \ub^\dagger D_\alpha \Ub \ub^\dagger \label{eq:Def-abbr-Dmix}
\end{align}
and $\Gamma_{\!\alpha}$ and $f^+_{\alpha\beta}$ evaluated at the classical solutions $\Ub$ and $\ub$. Hereby, the first flavour index $a$ in $\Dmix^{ab}$ denotes the vector-meson flavour and, hence, $a = 0, \dots, 8$ whereas the second flavour index $b$ denotes the pseudscalar flavour and $b = 1, \dots, 8$ as long as the $\eta$-singlet is not included (see section \ref{sec:etaprime} for inclusion of the $\eta$-singlet). However, the Gell-Mann matrix $\lambda^b$ corresponding to the pseudoscalar fluctuation only shows up in commutators such that including $b =0$ does not change the result and the summation rule \eqref{eq:GellMannSum} can be used.

Both $L$ and $J$ are of $\order(Q^2)$. In $K$, $\vartheta$ is of $\order(Q)$ and the remaining parts are of $\order(Q^3)$. 
To simplify finding possible ways of structuring terms, the calculation was additionally ordered in powers of $h_P$ yielding the following contributions to $\tr(\log \widetilde{D})_{\te{mix}}$:
\begin{align}
	&\! \! \! \! \!\!\tr(\log \widetilde{D})_{\te{mix}} = i \bar{\lambda}\!\! \int \df x \sum_{j=0}^{4} Q^{\te{mix}}_4(h_P^j) + \te{(finite)} + \order(Q^6) \allowdisplaybreaks \nn \\
	\bullet \,& Q^{\te{mix}}_4(h_P^0) = \frac{f_V^2}{16 \,F^2} \bra \zeta \sd (\zeta + 2\psi)^t - \psi \sd \psi^t \ket \allowdisplaybreaks \nn \\
	\bullet \,& Q^{\te{mix}}_4(h_P^1) \nn \\
	&= \frac{f_V^2 h_P}{16\, F^2}\, i \left\{ 2 \bra \delta \sd (\omega - \zeta)^t \ket + 4\bra (\psi^{\tau\eta}_\tau - \partial_\tau \omega^{\tau\eta} + d\omega^{\tau\eta}_\tau ) \vartheta^t_\eta \ket \nn \right. \adb \nn \\
	&\phantom{=} \, - 3 \bra \kappa^{\tau\eta}_\tau \vartheta^t_\eta \ket - \bra \vartheta^\tau G^\eta \zeta^t_{\tau\eta} \ket + \bra (\vartheta^\tau G^\eta + 2 \widetilde{G}^\eta \vartheta^\tau ) \omega^t_{\tau\eta} \ket \big\} \allowdisplaybreaks \nn \\
	\bullet \,& Q^{\te{mix}}_4(h_P^2) \nn \\
	&= \frac{f_V^2 h_P^2}{3 \sd 128 F^2} \big\{18 \mv^2 \bra \vartheta \sd \vartheta^t \ket + 18 \bra \mathcal{M} (\vartheta^t \sd \vartheta) \ket \nn \adb \\
	&\phantom{=} +  24 \bra \delta \sd \delta^t + 2 (\varphi - \partial \gamma + d\gamma)\sd \vartheta^t \ket - 36 \bra \iota \sd\vartheta^t \ket \nn \adb \\
	&\phantom{=} + 2 \bra 5 \Box \vartheta \sd \vartheta^t + 2 (\partial \vartheta) (\partial \vartheta^t) \ket + 24 \bra \delta^t_{\tau\eta} \vartheta^\tau G^\eta + \gamma^t_{\tau\eta} \widetilde{G}^\eta \vartheta^\tau \ket \nn \adb \\
	&\phantom{=} + 18 \bra F_\chi (\vartheta^t \sd \vartheta) \ket - 3 \bra \widetilde{E}^\tau_\tau (\vartheta \sd \vartheta^t) + 2 \widetilde{E}^{\tau\eta} \vartheta_\eta \vartheta_\tau \ket \nn \adb \\
	&\phantom{=} + 2 \bra 10 \widetilde{G}_\tau \partial^\eta \vartheta^\tau \vartheta_\eta^t - 8 \widetilde{G}^\tau (\partial \vartheta) \vartheta^t_\tau - 5 \widetilde{G}^\tau \partial_\tau \vartheta^\eta \vartheta_\eta^t \ket \nn \adb \\
	&\phantom{=} + 2 \bra -5 G^\tau \partial_\tau \vartheta_\eta^t \vartheta^\eta -2 (\vartheta \sd G) (\partial \vartheta^t) + 4 G^\tau \partial_\eta \vartheta_\tau^t \vartheta^\eta \ket \nn \adb \\
	&\phantom{=} + \bra -5 \widetilde{G}^\tau \widetilde{G}^\eta \vartheta_\tau \vartheta^t_\eta +(\vartheta^t \sd \widetilde{G}) ( \widetilde{G} \sd \vartheta) + \widetilde{G}^2 (\vartheta \sd \vartheta^t) \ket \nn \adb \\
	&\phantom{=} + \bra - 5 \widetilde{G}^\tau \vartheta^\eta G_\tau \vartheta^t_\eta + 2 (\widetilde{G} {\sd} \vartheta) (G \sd \vartheta^t) \ket - 5 \bra G^2 (\vartheta^t \sd \vartheta) \ket \nn \adb \\
	&\phantom{=} + \bra G^\tau G^\eta (\vartheta_\tau^t \vartheta_\eta + \vartheta^t_\eta \vartheta_\tau ) \ket \big\}	\allowdisplaybreaks  \adb \nn \\
	\bullet \,& Q^{\te{mix}}_4(h_P^3) = 0 \allowdisplaybreaks \nn \\
	\bullet \,& Q^{\te{mix}}_4(h_P^4) = \frac{f_V^4 h_P^4}{3 \sd 64\sd 128\, F^4} \bra 13 (\vartheta^t \sd \vartheta)^2 + (\vartheta \sd \vartheta^t)^2 + (\vartheta_\tau \vartheta^t_\eta)^2 \ket \label{eq:res-before-eom}
\end{align} 
Here, $\widetilde{G}$ and $\widetilde{E}$ denote $G$ and $E$, respectively, as given in \eqref{eq:DefEG} with $\widetilde{\Gamma}$ instead of $\Gamma$. Transposing a matrix refers only to transposing in flavour space. The matrix $F_\chi$ is part of the pseudoscalar contribution $\Dch$ and given by \cite{Gasser:1984gg} 
\begin{align}
	&F_\chi :=- \frac{1}{2} \,\partial_\tau G^\tau + \frac{1}{4} G^2 + \hat{\sigma} - \mathcal{M}, \nn \adb \\
	&\hat{\sigma}^{ab} := \frac{1}{2} \bra [ \lambda^a, \Umb_\tau ] [\lambda^b, \Umb^\tau ] \ket + \frac{1}{8} \bra \{ \lambda^a, \lambda^b \} ( \ub \chi^\dagger \ub + \ub^\dagger \chi \ub^\dagger ) \ket. \label{eq:Def-Fchi}
\end{align}

All terms except $Q_4^{\te{mix}}(h_P^2)$ can be calculated directly using the sum rules \eqref{eq:GellMannSum} and the trace relation \eqref{eq:RefDmu} yielding
\begin{align}
	\bullet \,& Q^{\te{mix}}_4(h_P^0) = - \frac{3 f_V^2}{8 F^2} \left( \bra F_L^2 + F_R^2 \ket - 2 \bra \bar{U}^\dagger F_R^{\mu\nu} \bar{U} F^L_{\mu\nu} \ket \right) \adb \nn \\
	\bullet \,& Q^{\te{mix}}_4(h_P^1) = - \frac{9f_V^2 h_P}{16 F^2} i \bra F_L^{\mu\nu} D_\mu \Ub^\dagger \, D_\nu \Ub + F_R^{\mu\nu} D_\mu \Ub \, D_\nu \Ub^\dagger \ket \adb \nn \\
	\bullet \,& Q^{\te{mix}}_4(h_P^4) = \frac{f_V^4 h_P^4}{64\sd 128 \, F^4} \Big\{ 24 \bra (D_\mu \Ub^\dagger \, D^\mu \Ub)^2 \ket \nn \\
	&\phantom{Q^{\te{mix}}_4(h) = } \ \ \  + 11 \left( 2 \bra D_\mu \Ub^\dagger \, D_\nu \Ub \ket^2 + \bra D_\mu \Ub^\dagger D^\mu \Ub \ket^2 \right) \! \Big\}.
\end{align} 
For calculating $Q_4^{\te{mix}}(h_P^2)$, the EOM of the LO $\chpt$ Lagrangian $\Lch^{\rm LO}$ is needed. It can be expressed as \cite{Bijnens:1999sh}
\begin{align}
	D_{\!\mu} \Umb^\mu = \frac{1}{4} \left( \chi^- - \frac{1}{3} \bra \chi^- \ket \right) \label{LOEOM}
\end{align}
with $D_{\!\mu} \Umb_\nu := \partial_\mu \Umb_\nu + [\Gamma_\mu, \Umb_\nu]$ and $\chi^- := \ub^\dagger \chi \ub^\dagger - \ub \chi^\dagger \ub$. Furthermore, the fields $f^{\pm}$ are equal to \cite{Bijnens:1999sh}
\begin{align}
	f^+_{\mu\nu} &= i \left( \Gamma_{\mu\nu} + \left[\Umb_\mu, \Umb_\nu \right] \right), \nn \\
	f^-_{\mu\nu} &= -i \left( D_{\!\mu} \Umb_\nu - D_{\!\nu} \Umb_\mu \right).
\end{align}
Therewith, the contribution proportional to $h_P^2$ can be rewritten as
\begin{align}
	&\! Q^{\te{mix}}_4(h_P^2) \nn  \\
	&= \frac{9 f_V^2 h_P^2}{32 F^2}\, \mv^2 \bra D_\mu \Ub^\dagger \, D^\mu \Ub \ket \adb \nn \\
	&\phantom{=} + \frac{f_V^2 h_P^2}{128 F^2} \left\{ \frac{1}{3} \left( 11 \bra D_\mu \Ub \, D^\mu \Ub^\dagger \ket^2 + 22 \bra D_\mu \Ub \, D_\nu \Ub^\dagger \ket ^2 \right. \right. \nn \adb \\
	&\phantom{=} \left.  - 6 \bra (D_\mu \Ub^\dagger \, D^\mu \Ub)^2 \ket \right) + 3 \bra \chi \bar{U}^\dagger + \bar{U} \chi^\dagger \ket \bra D_\mu \Ub^\dagger \, D^\mu \Ub \ket \nn \adb \\
	&\phantom{=} + 9 \bra (\chi^\dagger \bar{U} + \bar{U}^\dagger \chi) D_\mu \Ub^\dagger \, D^\mu \Ub \ket - \bra \chi^\dagger \bar{U} - \chi \bar{U}^\dagger \ket^2 \nn \adb \\
	&\phantom{=} + 3 \bra \chi^\dagger \bar{U} \chi^\dagger \Ub + \chi \Ub^\dagger \chi \Ub^\dagger \ket + 20 \bra \Ub^\dagger F_R^{\mu\nu} \Ub F_L^{\mu\nu} \ket \nn \adb \\
	&\phantom{=} + 28 \, i \bra F_R^{\mu\nu} D_\mu \Ub \, D_\nu \Ub^\dagger + F_L^{\mu\nu} D_\mu \Ub^\dagger \, D_\nu \Ub \ket \nn \adb \\
	&\phantom{=}  - 10 \bra F_R^{\mu\nu} F^R_{\mu\nu} + F_L^{\mu\nu} F^L_{\mu\nu} \ket - 6 \bra \chi^\dagger \chi \ket  \Big\} .
\end{align}
The contribution including $\Dmix$ renormalises the low-energy constant $F^2$ in the LO-$\chpt$ Lagrangian $\Lch^{\rm LO}$ and all constants except $L_6$ in the NLO Lagrangian $\Lch^{\rm NLO}$ (see subsection \ref{subsec:Renorm}).

\subsection{Renormalisation of the low-energy constants of the leading- and next-to-leading-order $\chpt$ Lagrangians} \label{subsec:Renorm}

At $\order(Q^4)$, the effective action is given by
\begin{align}
	& Z = \! \int \df x \Lch^{\te{cl}} + Z_{\te{one loop}} + \order(Q^6)
\end{align}
with $\Lch = \Lch^{\rm LO} + \Lch^{\rm NLO}$ as defined in Eq. \eqref{eq:ChPTLagr}. The one-loop infinities have to be absorbed by renormalising the low-energy constants ``const'' such that $Z$ is finite at $\order(Q^4)$ if expressed in terms of the renormalised low-energy constants $\te{(const)}^r$. We have the following low-energy constants at our disposal: $F$ and $B_0$ of the $Q^2$ Lagrangian $\Lch^{\rm LO}$ together with $L_1, \dotsc, L_{10}$, $H_1$ and $H_2$ of the $Q^4$ Lagrangian $\Lch^{\rm NLO}$.

\begin{figure}[h]
	{\centering
	\includegraphics[trim = 50 50 -50 50, width = 0.55\textwidth]{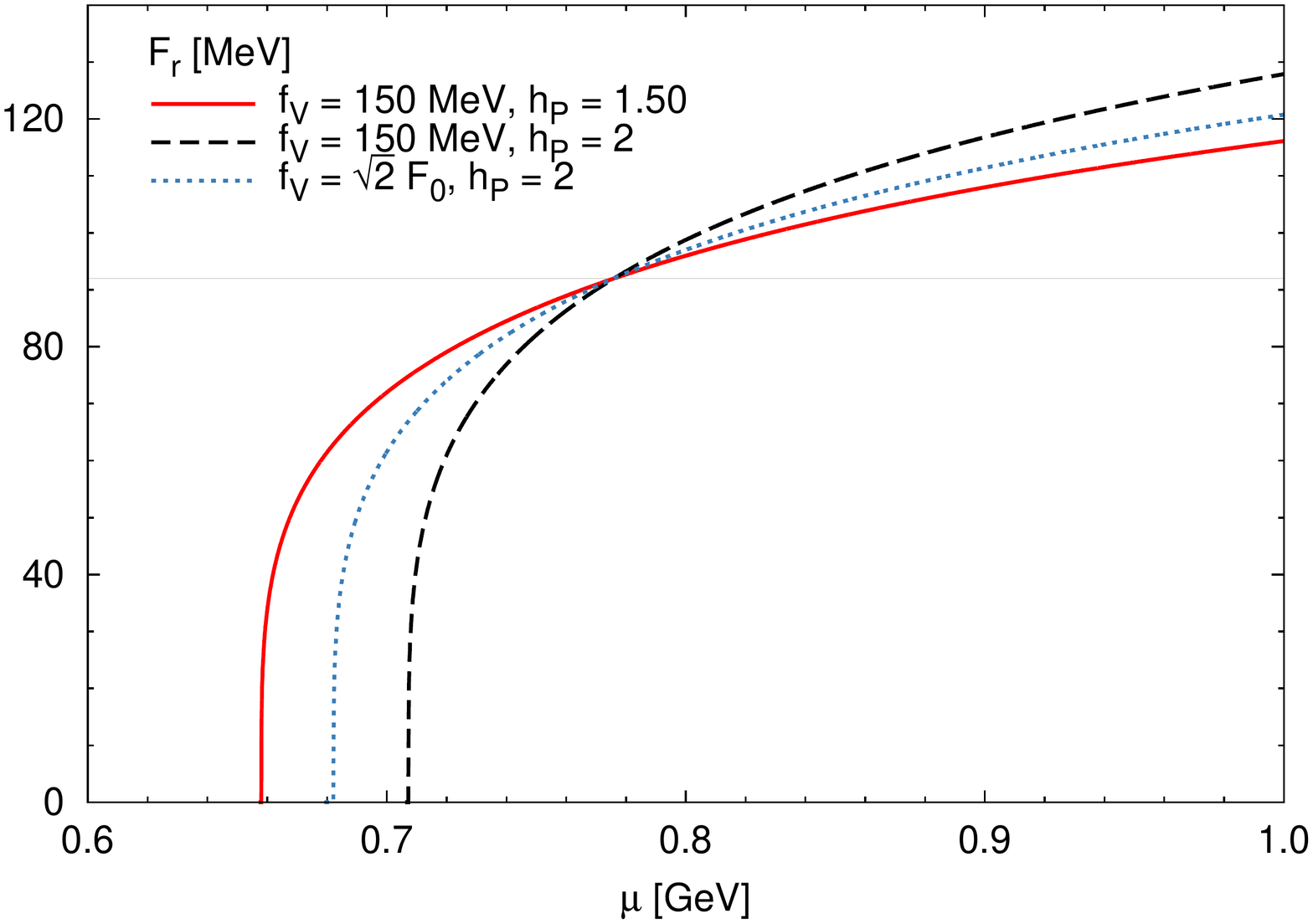}}
	\caption{Renormalised constant $F_r(\mu)$ as a function of the renormalisation scale $\mu$ (Eq.\ \eqref{eq:F2largeNc}) for different values of $f_V$ and $h_P$ (see legend).}
	\label{fig:MuDepF2}
\end{figure}

Only $Q_4^{\te{mix}}(h_P^2)$ is non-zero at $\order(Q^2)$ renormalising the wave-function renormalisation constant $F$ in $\Lch^{\rm LO}$ as
\begin{align}
	&F^2_r = F^2 + \frac{\varphi}{F^2_r} \left[ \bar{\lambda} - \frac{1}{16\pi^2} \left( \log \mu^2 + \te{finite} \right) \right] , \nn \adb \\
	 &\varphi := -\frac{9}{16} f_V^2 h_P^2 \mv^2 \label{eq:RenF}
\end{align}
depending on the renormalisation scale $\mu$ and for $\bar{\lambda}$ as defined in Eq.\ \eqref{eq:FormularMomInt}. In practice it is useful to expand $F^2_r$ in contributions sorted by the number of loops. Equivalently one can sort in inverse powers of the number of colours, $N_c$, assuming $N_c$ to be large. In this case, 
\begin{align*}
	\mv \in \order(1), \ F^2 \! \!, \, f_V^2 \in \order(N_c) \,. 
\end{align*}
Therewith, the dependence of $F^2_r$ on the renormalisation scale can be determined as
\begin{align}
	\frac{\te{d}F_r^2}{\te{d}\mu^2} = - \frac{\varphi}{16 \pi^2 \, F_r^2} \cdot \frac{1}{\mu^2} + \order(1/N_c). \nn
\end{align}
This differential equation can be solved for an arbitrary reference scale $\mu_0$ yielding
\begin{align}
	F_r^2 (\mu) := \sqrt{F_r^4(\mu_0) + \frac{2 \varphi}{16 \pi^2} \log \frac{\mu_0^2}{\mu^2}} + \order(1).  \label{eq:F2largeNc}
\end{align}
In Fig.\ \ref{fig:MuDepF2}, the renormalised constant $F_r(\mu)$ is plotted as a function of the scale $\mu$ assuming that the value $F {=} 92 \, \te{MeV}$ is reproduced for $\mu_0 {=} m_V {=} 0.776 \,\te{GeV}$. Hereby, two different values for both the parameter $h_P$ and the vector meson decay constant $f_V$ are used. On the one hand, $h_P$ has been determined from decays of light vector mesons into two pseudoscalar mesons in \cite{Lutz:2008km}\footnote{Note that the parameter $h_P$ was redefined compared to the definition used in \cite{Lutz:2008km}.}, $h_P {=} 1.50$. On the other hand, the KSFR relation $F_V {=} 2 G_V$ \cite{Ecker:1989yg} yielding $h_P {=} 2$ is used (see also Tab.\ \ref{tab:CompNotation}). The vector-meson decay constant is either approximated by $f_V {=} 150\,\te{MeV}$ \cite{Terschlusen:2013iqa} or by $f_V {=} \sqrt{2} F {=} \sqrt{2}\cdot 92 \,\te{MeV}$ \cite{Ecker:1989yg}. Note that $F_r(\mu)$ becomes imaginary for too small values of $\mu$. In general, Fig.\ \ref{fig:MuDepF2} displays a quite drastic renormalisation scale dependence of $F_r(\mu)$. Also the dependence on the actual values for the vector-meson coupling constants $h_P$ and $f_V$ is rather significant. To which extent all this carries over, for instance, to a vector-meson loop induced quark-mass dependence of the pseudoscalar decay constants remains to be seen \cite{MassPaper}; see the corresponding discussion in section \ref{sec:Intro}. 

The low-energy constants of $\Lch^{\rm NLO}$ are already renormalised by pure $\chpt$ \cite{Gasser:1984gg}\footnote{Note that the parameter $\bar{\lambda}$ is twice the corresponding parameter in \cite{Gasser:1984gg} yielding adapted coefficients in Tab.\ \ref{tab:VglRen}.},
\begin{align}
	&\left( L_i^r \right)_{\te{pure $\chpt$}} = \left( L_i \right)_{\te{pure $\chpt$}} + \frac{1}{2} \Gamma_i \bar{\lambda}, \nn \adb \\
	&\left( H_i^r \right)_{\te{pure $\chpt$}} = \left( H_i \right)_{\te{pure $\chpt$}} + \frac{1}{2} \Delta_i \bar{\lambda}, \nn \adb \\
	& \Gamma_1 = -\frac{3}{32}\, , \ \Gamma_2 = 2 \Gamma_1 \, , \ \Gamma_3 = 0 \, , \ \Gamma_4 = -\frac{1}{8}\, , \ \Gamma_5 = 3 \Gamma_4 \, , \nn \adb \\
	& \Gamma_6 = -\frac{11}{144}\, , \ \Gamma_7 = 0 \, , \ \Gamma_8 = -\frac{5}{48} \, , \ \Gamma_9 = -\frac{1}{4} \, , \nn \adb \\
	& \Gamma_{10} = - \Gamma_9 \, , \ \Delta_1 = - \Gamma_4 \, , \ \Delta_2 = 2 \Gamma_8 \,. \label{eq:RenParamChPT}
\end{align}
If loops with vector mesons are additionally taken into account, the renormalised constants will change to
\begin{align}
	&\left( L_i^r \right)_{\chpt + V} = \left( L_i \right)_{\chpt + V} + \left( \frac{1}{2} \Gamma_i + \Lambda_i \right) \! \bar{\lambda}, \nn \adb \\
	&\left( H_i^r \right)_{\chpt + V} = \left( H_i \right)_{\chpt + V} + \left( \frac{1}{2} \Delta_i + \eta_i \right) \! \bar{\lambda}, \nn \adb \\
	& \Lambda_1 = \frac{3}{64} -\frac{11}{6} h_P^2 \psi - \frac{11}{2} h_P^4 \psi^2 \,, \ \Lambda_2 = 2 \Lambda_1 \,, \nn \adb \\
	& \Lambda_3 = -\frac{9}{32} + h_P^2 \psi - 24 h_P^4 \psi^2 \,, \ \Lambda_4 = - \frac{3}{2} h_P^2 \psi \, , \nn \adb \\
	& \Lambda_5 = 3 \Lambda_4\,, \ \Lambda_6 = 0\,, \ \Lambda_7 = \frac{1}{2} h_P^2 \psi\,, \ \Lambda_8 = \Lambda_4\,, \nn \adb \\
	& \Lambda_9 = \frac{3}{8} + 2 h_P \psi \left( 7 h_P - 18 \right)\!, \ \Lambda_{10} = - \frac{3}{8} - 2 \psi \left( 5 h_P^2 + 24 \right)\!, \nn \adb \\
	& \eta_1 = -\frac{3}{16} + \psi \left(5 h_P^2 + 24 \right)\!, \ \eta_2 = -2 \Lambda_4  \label{eq:RenParam}
\end{align}
with $\psi := f_V^2/(128 \, F^2_r)$. At one-loop accuracy or in LO of a large-$N_c$ counting we have to make a choice for the value of $F_r$ to determine the numerical values for $\Lambda_i$ and $\eta_i$, respectively. We decided to use again $F^2_r(\mu {=} \mv) {=} ( 92\, \te{MeV})^2$ (cf.\ Eq.\ \eqref{eq:F2largeNc}). The other parameters $h_P$ and $f_V$ are varied as specified previously.

Comparing to the contributions from pure pseudoscalar loops \cite{Gasser:1984gg} shows that $L_6$ is only renormalised by loops emerging from the pure $\chpt$-Lagrangian while $L_3$ and $L_7$ are only renormalised by loops from Lagrangians containing vector mesons. Before looking at the numerical results we stress again that the divergence structure and the corresponding renormalisation scale dependence of the low-energy constants are not directly related to observables. Nonetheless a strong dependence might provide a first hint on possible momentum and/or quark-mass dependences of observables. Therefore we will determine how much the low-energy constants change numerically if the renormalisation point is varied within a reasonable range. We will compare this spread with the corresponding absolute size of the respective low-energy constant as determined from phenomenology. 

Before addressing this issue at the end of this section we want to highlight the opposite aspect, the fact that the low-energy constants are not observables. One result that points to this fact is the finding that the choice of different representations for the vector mesons leads to a different renormalisation of the low-energy constants. To display this issue we compare our results with the ones based on the hidden local gauge formalism (HLG) \cite{Harada:2003jx}. 

In Tab.\ \ref{tab:VglRen} we provide the numerical values for the renormalisation coefficients $\Gamma_i/2$ and $\Delta_i/2$ as generated by pure pseudoscalar loops and for $\Lambda_i$ and $\eta_i$ caused by loops including vector mesons. As one can see, the renormalisation coefficients are very sensitive to the actual choice of the parameters $h_P$ and $f_V$. Whenever non-vanishing the renormalisation coefficients from pure pseudoscalar loops and from loops including vector mesons are comparable in absolute size except for the quantities $L_{10}$ and $H_1$. We have not found a deeper reason for this fact, but we note that these are the quantities that contain two field-strength tensors of the external vector and axial-vector sources. In HLG a much larger renormalisation effect can be observed for some of the low-energy constants. This stresses again the representation dependence of the results for non-observable quantities like the low-energy constants. If these differences have any impact on observables remains to be seen. 
\begin{table}[h]
\caption{Numerical values for the renormalisation coefficients in different frameworks. The first column shows the respective low-energy constant. The second to fourth column provide the renormalisation coefficients generated by loops including vector mesons as given in Eq.\ \eqref{eq:RenF} and Eq.\ \eqref{eq:RenParam}. For instance, the value of $\Lambda_1$ is given in the row of $L_1$. The values for the parameters $h_P$, $f_V$ and $F_r$ are discussed in the main text. The fifth column yields the corresponding HLG value. The last column provides the $\chpt$ result \cite{Gasser:1984gg} for the renormalisation coefficients generated by loops that only contain pseudoscalar mesons.}
\label{tab:VglRen}
\begin{tabular}{l||c c|c||c||c}
	 & \multicolumn{3}{c||}{loops incl. vector mesons} & \multirow{3}{*}{HLG } & \multirow{3}{2em}{pure $\chpt$} \\
	 & \multicolumn{2}{l|}{$f_V {=} 150\,\te{MeV}$} & \multirow{2}{4.5em}{$f_V{=} \sqrt{2}F_0$ $h_P{=}2$}  &  & \\
	 & $h_P {=}1.50$ & $h_P {=} 2$ &  &  & \\ \hline \hline
	 & & & & & \\[-0.75em]
	$F$ & $-0.017$ & $-0.030$ & $-0.023$ & $\phantom{-}2.805$ & $\phantom{-.000}0$ \\ \hline
	 & & & & & \\[-0.75em]
	$L_1$ & $-0.051$ & $-0.143$ & $-0.089$  & $-0.060$ & $-0.047$ \\ 
	$L_2$ & $-0.102$ & $-0.287$ & $-0.178$ & $-0.120$ & $-0.094$ \\ 
	$L_3$ & $-0.287$ & $-0.364$ & $-0.313$ & $\phantom{-}2.267$ & $\phantom{-.000}0$ \\
	$L_4$ & $-0.070$ & $-0.125$ & $-0.094$ & $\phantom{-}0.160$ & $-0.063$ \\
	$L_5$ & $-0.210$ & $-0.374$ & $-0.281$ & $\phantom{-}0.479$ & $-0.188$ \\
	$L_6$ & $\phantom{-.000}0$ & $\phantom{-.000}0$ & $\phantom{-.000}0$ & $\phantom{-}0.115$ 	& $-0.038$ \\
	$L_7$ & $\phantom{-}0.023$ & $\phantom{-}0.042$ & $\phantom{-}0.031$ & $-0.032$ & $\phantom{-.000}0$ \\
	$L_8$ & $-0.070$ & $-0.125$ & $-0.094$ & $\phantom{-}0.149$ & $-0.052$  \\
	$L_9$ & $-0.092$ & $\phantom{-}0.043$  & $\phantom{-}0.125$ & $-2.371$ & $-0.125$ \\
	$L_{10}$ & $-1.839$ & $-2.203$  & $-1.750$ & $\phantom{-}2.371$ & $\phantom{-}0.125$ \\ \hline
	 & & & & & \\[-0.75em]
	$H_1$ & $\phantom{-}0.545$ & $\phantom{-}0.726$ & $\phantom{-}0.500$ & $\phantom{-}1.315$ & $\phantom{-}0.063$ \\
$H_2$ & $\phantom{-}0.140$ & $\phantom{-}0.249$ & $\phantom{-}0.188$ & $-0.090$ & $-0.104$ 
\end{tabular}
\end{table} 

Finally we introduce the renormalisation-scale dependence (variation) of the NLO low-energy constants by
\begin{align}
  \Delta L_i & := \left[ L_i^r(\mu_2)- L_i^r(\mu_1)\right]_{\chpt+V} \nonumber \\
  & = - \frac{1}{16\pi^2} \left( \frac{1}{2} \Gamma_i + \Lambda_i \right) \log \frac{\mu_2^2}{\mu_1^2} 
  \label{eq:MuDepLi}
\end{align}
for two scales $\mu_1$ and $\mu_2$. In Tab.\ \ref{tab:VglMuDep}, the changes in the low-energy constants $L_1, \ldots, L_{10}$ for $\mu_1 {=} 0.5 \, \te{GeV}$ and $\mu_2 {=} 1 \, \te{GeV}$ for a calculation with both pseudoscalar and vector mesons in the loop, for a pure $\chpt$ calculation, and for a calculation using the HLG formalism \cite{Harada:2003jx} are compared to the phenomenologically determined values for the low-energy constants based on pure $\chpt$ \cite{Gasser:1984gg, Bijnens:2014lea}. We observe that the changes caused by varying the renormalisation scale are comparable in size to the absolute values of the low-energy constants. 
\begin{table}[h]
\caption{Variation of the low-energy constants with the renormalisation scale as given in Eq.\ \eqref{eq:MuDepLi} including loops with vector mesons or using the HLG formalism \cite{Harada:2003jx} or including only pseudoscalar loops (``pure $\chpt$''), respectively. The used renormalisation points are $\mu_1 {=} 0.5 \, \te{GeV}$ and $\mu_2 {=} 1 \, \te{GeV}$. The results are compared to the phenomenologically determined values for the low-energy constants \cite{Gasser:1984gg, Bijnens:2014lea}. All values are given in units of $10^{-3}$. }
\label{tab:VglMuDep}
\begin{tabular}{l||c c|c||c||c||c}
	 & \multicolumn{5}{c||}{renormalisation-point variation} & \multirow{3}{3.8em}{\centering{phenom. value for $L_i$}} \\[0.1em]
	 & \multicolumn{2}{l|}{$f_V {=} 150\,\te{MeV}$} & \multirow{2}{4.5em}{$f_V{=} \sqrt{2}F_0$ $h_P{=}2$}  & \multirow{2}{*}{HLG } & \multirow{2}{2em}{pure $\chpt$} &  \\
	 & $h_P {=}1.50$ & $h_P {=} 2$ &  &  & & \\ \hline \hline
	 & & & & & & \\[-0.75em]
	$\Delta L_1$ & $\phantom{1}0.9$ & $\phantom{1}1.7$ & $\phantom{1}1.2$ & $\phantom{-}\ \, 0.5$ & $\phantom{-}0.4$ & $\phantom{-}1.0 \pm 0.1$ \\ 
	$\Delta L_2$ & $\phantom{1}1.7$ & $\phantom{1}3.3$ & $\phantom{1}2.4$ & $\phantom{-}\ \, 1.1$ & $\phantom{-}0.8$ & $\phantom{-}1.6 \pm 0.2$ \\  
	$\Delta L_3$ & $\phantom{1}2.5$ & $\phantom{1}3.2$ & $\phantom{1}2.7$ & $-19.9$ & $\phantom{-0.}0$ & $-3.8 \pm 0.3$ \\
	$\Delta L_4$ & $\phantom{1}1.2$ & $\phantom{1}1.6$ & $\phantom{1}1.4$ & $\ \, -1.4$ & $\phantom{-}0.6$ & $\phantom{-}0.0 \pm 0.3$ \\
	$\Delta L_5$ & $\phantom{1}3.5$ & $\phantom{1}4.9$ & $\phantom{1}4.1$ & $\ \, -4.2$ & $\phantom{-}1.7$ & $\phantom{-}1.2 \pm 0.1$ \\ 
	$\Delta L_6$ & $\phantom{1}0.3$ & $\phantom{1}0.3$ & $\phantom{1}0.3$ & $\ \, -1.0$ & $\phantom{-}0.3$ & $\phantom{-}0.0 \pm 0.4$ \\
	$\Delta L_7$ & $\!\!-0.2$ & $\!\!-0.4$ & $\!\!-0.3$ & $\phantom{-}\ \, 0.3$ & $\phantom{-0.}0$ & $-0.3 \pm 0.2$ \\
	$\Delta L_8$ & $\phantom{1}1.1$ & $\phantom{1}1.6$ & $\phantom{1}1.3$ & $\ \, -1.3$ & $\phantom{-}0.5$ & $\phantom{-}0.5 \pm 0.2$ \\
	$\Delta L_9$ & $\phantom{1}1.9$ & $\phantom{1}0.7$ & $\phantom{1}0.0$ & $\phantom{-}20.8$ & $\phantom{-}1.1$ & $\phantom{-}6.9 \pm 0.7$ \\
	$\Delta L_{10}$ & $15.1$ & $18.2$ & $14.3$ & $-20.8$ & $-1.1$ & $-5.5 \pm 0.7$ 
\end{tabular}
\end{table} 

\section{One-loop contributions up to $\order(Q^4)$ including the $\eta$-singlet} \label{sec:etaprime}

In the calculations presented so far, the Goldstone-boson octet described by the matrix $\Phi$ (cf.\ Eq.\ \eqref{eq:Def-pseudo-octet}) was used \textit{i.e.}, the physical $\eta$-meson was approximated by the (unphysical) octet state $\eta_8$. If the $\etapr$-meson is included additionally, the Goldstone-boson nonet with the singlet state $\phi_0$ has to be considered, \textit{i.e.},
\begin{align}
	&\Phi \mapsto \Phi_{\te{octet}} + \sqrt{\frac{2}{3}}\, \phi_0 \, \mathbb{1} \,.
\end{align}

A formally systematic framework for the low-energy effective theory of the pseudoscalar nonet is $\chpt$ for a large number of colours \cite{Kaiser:2000gs}. There the LO large-$N_c$ $\chpt$ Lagrangian is given by\footnote{For the present work we ignore the vacuum angle $\theta$ that is related to the chiral anomaly and to strong P and CP violation \cite{Bailin-Love}.}
\begin{align}
  &\begin{array}{c|c c} \Lag_{\rm LO}^{+ \etapr} = \Lch^{\rm LO} & & \hspace{-0.75em} -\, \frac12 \, m_0^2 \phi_0^2 \,, \\[-0.5em] 
    & \scriptstyle{\Phi = \sum\limits_{c=0}^8 \lambda^c \phi_c} &
  \end{array} 
  \label{eq:lnclo}
\end{align}
with $m_0^2 = 6\tau/F^2$ and the topological susceptibility $\tau$. 

In the power counting of large-$N_c$ $\chpt$ the ``NLO'' Lagrangian in \eqref{eq:ChPTLagr} contains NLO terms and 
terms of next-to-next-to-leading order (N$^2$LO). In addition, the NLO Lagrangian of large-$N_c$ $\chpt$ receives additional 
contributions \cite{Kaiser:2000gs}. To cancel the infinities of one-loop diagrams including vector mesons we will need parts of 
the LO, NLO and N$^2$LO Lagrangians of large-$N_c$ $\chpt$. Instead of writing down all these Lagrangians we restrict ourselves 
to the terms that are needed for the renormalisation of the loops including vector mesons. 
These terms are covered by \eqref{eq:ChPTLagr}, \eqref{eq:lnclo} and 
\begin{align}
  &\begin{array}{c|c c} \Lag_{\rm ct}^{+ \etapr} = \Lch^{\rm NLO} & & \hspace{-0.75em} 
    +\, \frac{1}{2 \sqrt{6}} \, F \, \tilde \Lambda_2 \, i \phi_0 \, \bra \chi^\dagger U -\chi U^\dagger \ket \,. \\[-0.5em] 
    & \scriptstyle{\Phi = \sum\limits_{c=0}^8 \lambda^c \phi_c} &
  \end{array} 
  \label{eq:lncnlo}
\end{align}

In the following we focus on the changes caused by this extension of the framework. We provide less details since most of the calculations technically proceed in the very same way. 
The matrix $\Dch$ corresponding to the extended LO Lagrangian \eqref{eq:lnclo} changes to 
\begin{align}
  &\begin{array}{c|c c} [\Dch^{+ \etapr}]^{ab} = [\Dch]^{ab} & & \hspace{-0.75em} +\, 2 \, m_0^2 \delta^{a0} \delta^{b0} \,. \\[-0.5em]
    &  {\scriptstyle{\Phi = \sum\limits_{c=0}^8 \lambda^c \phi_c}} &
  \end{array}
\end{align}
In contrast to the case for the Goldstone-boson octet, the matrix $\Dch^{+\etapr}$ is not diagonal in the limit of no external fields, 
\begin{align}
	\bullet \ & [\Dch^{0, +\etapr}(k)]^{-1}_{00} = \alpha^+ (k^2 - M^+)^{-1} + \alpha^- (k^2 - M^-)^{-1}, \nn \adb \\
	\bullet \ & [\Dch^{0, +\etapr}(k)]^{-1}_{11 \, / \ldots / \, 33}=[ 2F^2  (k^2 - \frac{1}{2} M_\pi^2 ) ]^{-1}, \nn \adb \\
	\bullet \ & [\Dch^{0, +\etapr}(k)]^{-1}_{44 \, / \ldots / \, 77} = [ 2 F^2 (k^2 - \frac{1}{2} M_K^2 ) ]^{-1}, \nn \adb \\
	\bullet \ & [\Dch^{0, +\etapr}(k)]^{-1}_{08} = [\Dch^{0, +\etapr}(k)]^{-1}_{80} \nn \\
	& \phantom{[\Dch^{0, +\etapr}(k)]^{-1}_{08}} \sim \left[4 F^4 (k^2 - M^+)(k^2 - M^-)\right]^{-1}, \nn \adb \\
	\bullet \ & \text{all other matrix entries equal to zero.}
	\label{eq:D0mitEtapr}
\end{align}
Hereby, $M^\pm$ and $\alpha^\pm$ denote a combination of pion mass $M_\pi$, kaon mass $M_K$ and topological susceptibility $\tau$ with $\alpha^+ {+} \alpha^- {=} 1$. In \cite{Kaiser:2000gs}, the changes in the renormalisation of the low-energy constants caused by adding the $\eta$-singlet to the pure $\chpt$-Lagrangian have been determined. They are not repeated in this article.

For the calculation of $\tr(\log D)$ containing $\Dmix$ including the $\eta$-singlet, note that $\widetilde{\Delta}_{a0} = 0$ for all vector-meson flavors $a = 0, \dotsc, 8$ as already mentioned in subsection \ref{subsec:mix}. Furthermore, the non-zero terms including $[\Dch^{0, +\etapr}(k)]^{-1}_{08}$ or $[\Dch^{0, +\etapr}(k)]^{-1}_{80}$ are proportional to
\begin{align}
	\int \frac{\! \! d^{4+2\eps}k}{(2\pi)^{4+2\eps}} \, k^2 (k^2-m^2)^{-4} = \te{(finite)}.
\end{align}
Additionally, if using $\alpha^+ {+} \alpha^- {=} 1$ the only difference for $\tr(\log \widetilde{D})$ containing $\Dmix$ with and without the $\eta$-singlet is visible in terms containing the pseudoscalar masses explicitly. For calculations without the $\eta$-singlet, the only terms containing these masses explicitly are
\begin{align}
	Q_4^{\te{mix}}(h_P^2) \sim \bra \mathcal{M} (\vartheta^t \sd \vartheta) \ket + \bra \te{(mass part of }F_\chi \te{)} (\vartheta^t \sd \vartheta) \ket
\end{align}
with $\vartheta$ and $F_\chi$ as defined in Eq.\ \eqref{eq:Def-abbr-Dmix} and \eqref{eq:Def-Fchi}, respectively. However, inserting $\Dch^{+\etapr}$ into the equations for those terms yields the same result with the mass matrix $\mathcal{M}$ and the corresponding part in $F_\chi$ modified according to \eqref{eq:D0mitEtapr}. Therefore, the sum of these two terms vanishes both for the calculation with and without the $\eta$-singlet. Hence, the parts containing $\Dmix$ in $\tr(\log \widetilde{D})$ are the same up to finite parts and terms of $\order(Q^6)$ for both not including and including the $\eta$-singlet. Thus we are back to the same expression as given in (\ref{eq:res-before-eom}). 

However, the final results of subsection \ref{subsec:mix} have been obtained by using the EOM (\ref{LOEOM}) emerging from the LO Lagrangian. In the presence of the singlet field and in particular due to the effect from the topological susceptibility the EOM changes to 
\begin{align}
  D_{\!\mu} \Um^\mu = \frac{1}{4} \chi^- - i \, \frac{m_0^2}{\sqrt{6} F} \, \phi_0   \,.
  \label{LOEOM-largenc}
\end{align}
The results from the previous section are modified and extended in the following way: The results for the renormalisation of all the previously introduced low-energy constants remains the same except for $L_7$ which now does {\em not} receive any renormalisation. In addition, the new low-energy constants $\tau$ and $\tilde \Lambda_2$ receive the following renormalisation from loops with vector mesons:
\begin{eqnarray}
  \tau_r & = & \tau - \frac{9 f_V^2 h_P^2 \tau_r^2}{8 F_r^6} \, \bar\lambda \,, \nonumber \\
  \tilde \Lambda_2^r & = & \tilde \Lambda_2 - \frac{9 f_V^2 h_P^2 \tau_r}{8 F_r^6} \, \bar\lambda  \,.
  \label{eq:taulambda2}
\end{eqnarray}
As already spelled out, everything else remains unchanged.

\section{Outlook}
In the present work, the infinity structure and corresponding renormalisation-scale dependence of all $\chpt$-low-energy constants up to chiral order $Q^4$ have been determined. Thereby, the finite parts of the loops with vector mesons depend in addition on the masses of vector and pseudoscalar mesons and on the external momenta. It is therefore interesting how physical observables depend on these. In the follow-up work \cite{MassPaper} we will study the influence of loops with vector mesons on the pseudoscalar properties (mass and decay constant) within the same framework as used in the present work. 

Furthermore, a plausibility check of the counting scheme with both light pseudoscalar and vector mesons as degrees of freedom as suggested in \cite{Terschlusen:2012xw} can be performed. Therein, vector mesons are counted as soft, \textit{i.e.}, the vector meson mass $\mv$ is of chiral order $Q$. Therefore, one-loop diagrams of $\order(Q^4)$ could have a chiral structure of $\order(Q^6)$ divided by $\mv^2$. Since the corresponding infinities would have no counter terms in the NLO-$\chpt$ Lagrangian of $\order(Q^4)$, all these infinities either have to vanish directly or for specific parameter combinations within a reasonable framework. Note that such a plausibility check has to be performed for the full Lagrangian given in \cite{Terschlusen:2012xw} and not only for the restricted Lagrangian as used in this work. Additionally, calculations with vector mesons as non-vanishing classical fields and the renormalisation of parameters in an NLO Lagrangian with vector mesons is of interest.

\begin{appendix}
\section{Heat-kernel calculation for a toy model with only one charged vector meson} \label{app:heat-kernel}

In subsection \ref{subsec:Heat-kernel}, the heat-kernel approach is discussed and that it is not applicable to Lagrangians with vector mesons represented by antisymmetric tensor fields. Here, the heat-kernel approach is tried to be applied to a vector-meson Lagrangian. It is discussed in greater detail why such a procedure is not applicable. For that, consider a toy Lagrangian for one complex vector-meson flavour, $V^{\mu\nu} = - V^{\nu\mu}$,
\begin{align*}
	\mathcal{L}_{\te{toy}} =  - \left( D_\mu V^{\mu\nu} \right)^\dagger D^\rho V_{\rho\nu} + m^2 V^\dagger_{\mu\nu} V^{\mu\nu}
\end{align*}
with $D_\mu = \partial_\mu + i \Gamma_\mu$ for an arbitrary $\Gamma_\mu = \Gamma_\mu^\dagger$. The vector field is split into its projections,
\begin{align*}
	V^{\mu\nu}(x)  &= \sum_{j=A,V} \int \df y  \int \frac{\te{d}^4k}{(2\pi)^4} e^{-ik(x-y)} \widetilde{P}_j^{\mu\nu\alpha\beta}(k) V_{\alpha\beta}(y) \\
	&=: \mathcal{A}^{\mu\nu}(x) + \mathcal{V}^{\mu\nu}(x)\,
\end{align*}
including the antisymmetric projection operators in momentum space as defined in Eq.\ \eqref{eq:Def:proj-operators}. Therewith, the Lagrangian can be rewritten as
\begin{align*}
	\mathcal{L}_{\te{toy}} =& \  \Vpr_{\mu\nu}^\dagger \Box \Vpr^{\mu\nu} + m^2 \left( \Apr^\dagger_{\mu\nu} \Apr^{\mu\nu} + \Vpr^\dagger_{\mu\nu} \Vpr^{\mu\nu} \right)  \adb \\
	& + i \left[ \Gamma^\mu \left(\Apr+\Vpr\right)_{\mu\nu}^\dagger \, \partial^\rho \Vpr_{\rho\nu} - \partial^\mu \Vpr_{\mu\nu}^\dagger \, \Gamma_\rho \left(\Apr+\Vpr\right)^{\rho\nu} \right] \adb \\
	&- \Gamma^\mu \Gamma_\rho \, \left(\Apr+\Vpr\right)^\dagger_{\mu\nu} \left(\Apr+\Vpr\right)^{\rho\nu}\,.
\end{align*}
In this notation, one can identify a non-propagating mode $\Apr$, \textit{i.e.}, a field with mass term only and without kinetic term, and a propagating mode $\Vpr$, \textit{i.e.}, a field with both mass and kinetic term. First, the equation of motion (EOM) for the classical non-propagating mode $\bar{\Apr}$ has to be calculated yielding
\begin{align*}
	0 = m^2 \bar{\Apr}^{\mu\nu} - P_1^{\mu\nu\alpha\beta} \Gamma_\alpha \Gamma^\rho \left( \bar{\Apr} + \Vpr \right)_{\rho\beta} + i P_1^{\mu\nu\alpha\beta} \Gamma_\alpha \, \partial^\rho \Vpr_{\rho\beta}.
\end{align*}
Note that this EOM depends on the full field $\Vpr$ not only on its classical part $\bar{\Vpr}$. The Lagrangian can be expanded around $\bar{\Apr}$ via $\Apr =: \bar{\Apr} + \delta\Apr$ yielding
\begin{align*}
	\mathcal{L}_{\te{toy}} =& \  \Vpr_{\mu\nu}^\dagger \Box \Vpr^{\mu\nu} + m^2 \left( \bar{\Apr}^\dagger_{\mu\nu} \bar{\Apr}^{\mu\nu} + \Vpr^\dagger_{\mu\nu} \Vpr^{\mu\nu} \right)  \adb \\
	& + i \left[ \Gamma^\mu \left(\bar{\Apr}+\Vpr\right)_{\mu\nu}^\dagger \, \partial^\rho \Vpr_{\rho\nu} - \partial^\mu \Vpr_{\mu\nu}^\dagger \, \Gamma_\rho \left(\bar{\Apr}+\Vpr\right)^{\rho\nu} \right] \adb \\
	&- \Gamma^\mu \Gamma_\rho \, \left(\bar{\Apr}+\Vpr\right)^\dagger_{\mu\nu} \left(\bar{\Apr}+\Vpr\right)^{\rho\nu} + \# \left( \delta\Apr \right)^2 + \order(\delta\Apr^3)\,.
\end{align*}
In principle, a heat-kernel calculation for the term quadratic in $\delta\Apr$ has to be performed yet this  yields zero in dimensional regularisation anyway. Next, the EOM for the classical propagating mode $\bar{\Vpr}$ is determined, 
\begin{align*}
	0 =&\  \Box \bar{\Vpr}^{\mu\nu} + m^2 \bar{\Vpr}^{\mu\nu} - P_1^{\mu\nu\alpha\beta} \Gamma_\alpha \Gamma^\rho \left( \bar{\Vpr} + \bar{\Apr}^0 \right)_{\rho\beta} \\
	& + i P_1^{\mu\nu\alpha\beta} \left[ \Gamma_{\alpha} \, \partial^\rho \bar{\Vpr}_{\rho\beta} + \partial_\alpha \left( \Gamma^\rho \bar{\Vpr}_{\rho\beta} + \Gamma^\rho \bar{\Apr}^0_{\rho\beta}  \right) \right] 
\end{align*} 
with $\bar{\Apr}^0 := \bar{\Apr}(\bar{\Vpr}) =: \bar{\Apr}(\Vpr) {-} a$, \textit{i.e} the classical non-propa{-}gating mode evaluated at the classical propagating mode. Recall that in the last formulation of the Lagrangian $\bar{\Apr}$ was used not $\bar{\Apr}^0$. The Lagrangian can be written in terms of $\Vpr = \bar{\Vpr} + \delta\Vpr + \order(\delta\Vpr)^2$ as
\begin{align*}
	\mathcal{L}_{\te{toy}} =& \ (\delta\Vpr)^\dagger_{\mu\nu} \Box (\delta\Vpr)^{\mu\nu} + m^2 \left( \delta\Vpr - a \right)^\dagger_{\mu\nu} \left( \delta\Vpr - a \right)^{\mu\nu} \\
	& + i \left[ \Gamma^\mu (\delta\Vpr)^\dagger_{\mu\nu} \, \partial_\rho (\delta\Vpr)^{\rho\nu} - \partial^\mu (\delta\Vpr)^\dagger_{\mu\nu} \, \Gamma_\rho (\delta\Vpr)^{\rho\nu} \right] \\
	& + \Gamma^\mu \Gamma_\rho \left[ (\delta\Vpr)^\dagger_{\mu\nu} \, (\delta\Vpr)^{\rho\nu} - a^\dagger_{\mu\nu} a^{\rho\nu} \right] + \# \left( \delta\Apr \right)^2 \\
	& + \te{(terms with $\bar{\Apr}^0$, $\bar{\Vpr}$ only)} + \order(\delta^3)\,.
\end{align*}
From the EOM for $\bar{\Apr}$ the field $a$ can be determined as a function of $\delta\Vpr$, 
\begin{align*}
	&a_{\mu\nu} = J^{-1}_{\mu\nu\alpha\beta} \left[ m^2 (\delta\Vpr)^{\alpha\beta} - i P_1^{\alpha\beta\rho\sigma} \Gamma_\rho \, \partial^\tau (\delta\Vpr)_{\tau\sigma} \right] - \delta\Vpr_{\mu\nu} \, , \adb \\
	& J^{\mu\nu\alpha\beta} := \left(m^2 g_\rho{}^{\alpha}  - \Gamma_ \rho \Gamma^\alpha  \right) P_1^{\mu\nu\rho\beta}\,.
\end{align*}
Therewith, the relevant part in the Lagrangian, \textit{i.e.}, the terms proportional to $(\delta\Vpr)^2$, can be identified,
\begin{align*}
	\mathcal{L}_{\te{toy}} =& \  (\delta\Vpr)^\dagger_{\mu\nu} \Box (\delta\Vpr)^{\mu\nu} - 2 m^2 (\delta\Vpr)^\dagger_{\mu\nu} (\delta\Vpr)^{\mu\nu} \adb \\
	& + 2 i \left[ \partial^\mu (\delta\Vpr)^\dagger_{\mu\nu} \, \Gamma^\rho (\delta\Vpr)_{\rho\nu} - \Gamma^\mu (\delta\Vpr)_{\mu\nu}^\dagger \, \partial^\rho (\delta\Vpr)_{\rho\nu} \right] \adb \\
	& - (J^{-1})^{\mu\nu\alpha\beta} \, \Gamma_\mu \partial^\rho (\delta\Vpr)^\dagger_{\rho\nu} \, \Gamma_\alpha \partial^\sigma (\delta\Vpr)_{\sigma\beta} \adb \\
	& + m^4 (\delta\Vpr)^\dagger_{\mu\nu} \,  (J^{-1})^{\mu\nu\alpha\beta} \, (\delta\Vpr)_{\alpha\beta} + \# \left( \delta\Apr \right)^2 \adb \\
	& + \te{(terms with $\bar{\Apr}^0$, $\bar{\Vpr}$ only)} + \order(\delta^3)\,.
\end{align*}
Since the Lagrangian has to be antisymmetric in the Lorentz indices, it can be rewritten in the form 
\begin{align*}
	\mathcal{L}_{\te{toy}} &= (\delta\Vpr)^\dagger_{\mu\nu} \left[ (2m^2 + R) + S_\tau \partial^\tau + \left( 1 - T \right) \Box \right]^{\mu\nu\alpha\beta} (\delta\Vpr)_{\alpha\beta}  \adb \\
	& \phantom{=} \ + \ldots \adb \\
	&=: (\delta\Vpr)^\dagger_{\mu\nu}\,  D^{\mu\nu\alpha\beta} (\delta\Vpr)_{\alpha\beta} + \ldots
\end{align*}
with symmetric matrices $R$ and $T$ and an antisymmetric matrix $S$. Furthermore, $R$, $S$ and $T$ will vanish in the limit of no external fields.

In a heat-kernel approach, the matrix element for $D$ is given by
\begin{align*}
	&\bra x | e^{-\lambda D} | y \ket = (4 \pi \lambda)^{-d/2} \exp{\left[\frac{(x-y)^2}{4 \lambda} \, - 2 m^2 \lambda \right]} H(x|\lambda|y), \adb \\
	& H(x| \lambda |y ) = \sum_{n=0}^{\infty} \lambda^n H_n(x|y)\,.
\end{align*}
The differential equation determined from this matrix element, 
\begin{align*}
	\frac{\partial}{\partial\lambda} \bra x \left| e^{-\lambda D} \right| y \ket  = - D_x \bra x \left| e^{-\lambda D} \right| y \ket,
\end{align*}
can be written in powers of $\lambda$. Since $\lambda$ is arbitrary, each order of $\lambda$ yields a recursive equation for the $H_n$ which can be used to determine $H_2$\footnote{Recall from subsection \ref{subsec:Heat-kernel} that only $H_2$ is needed to identify the infinities for $d{=}4$ dimensions.}. In particular, the contribution proportional to $\lambda^{-2}$ yields
\begin{align*}
	0 = \frac{1}{4} (x-y)^2 \, T^{\mu\nu\alpha\beta} H_0(x|y)_{\alpha\beta\rho\sigma} .
\end{align*}
Since $T \neq 0$ by definition and $(x,y)$ are arbitrary, 
\begin{align*}
	H_0 (x|y) = 0\,.
\end{align*}
However, this result is contrary to the initial condition 
\begin{align*}
	H_0 (x|y) = 1 + \order(x-y). 
\end{align*}
Therefore, the heat-kernel approach is not applicable for the toy Lagrangian $\mathcal{L}_{\te{toy}}$ for one vector meson. The same procedure can be applied to the full Lagrangian $\Lvec$ for vector mesons yielding that a heat-kernel approach is not applicable.

\section{Example of an integral for one-loop contributions in powers of $D {-}D_0$} \label{app:Ex-int}

The general procedure how to calculate the one-loop contribution in powers of $\delta D {=} D {-} D_0$ is described in subsection \ref{subsec:SumN}. In this section, an integral contributing to the second term in the sum \eqref{eq:SumN} for $\tr(\log D)$ is determined as an example for such an calculation. Thereby, a contribution to $\tr(\log \Dmix)$ is chosen, 
\begin{align*}
	I :=& \,\frac{i f_V^2 h_P}{4} \int \df x \te{d}^4\! y \te{d}^4\! x' \te{d}^4\! y' \int \frac{\te{d}^4\! k \te{d}^4\! p}{(2\pi)^8} \, e^{ik(x-x')+ip(y-y')} \\
	& \ \cdot \bra P [\Dvec^0(k)]^{-1}_{\mu\nu\alpha\beta} \, \zeta^{\alpha\beta}(x') \, \delta(x'-y) \, [\Dch^0(p)]^{-1} \right. \\
	& \ \phantom{\cdot \bra \right.} \left. \cdot  P_1^{\mu\nu\rho\sigma} g_{\rho\eta} \vartheta_\sigma(x) \,  \partial^\eta_x \delta(x-y') \ket \\
	\in& \,\tr \left[ ( D_0^{-1} \delta D)^2 \right] 
\end{align*} 
for $P \Dvec^0$, $\zeta$ and $\vartheta$ as defined in subsection \ref{subsec:vec} and \ref{subsec:mix}, respectively. After partial integration with respect to $x$, both $\delta$-functions in coordinate space can be evaluated,
\begin{align*}
	I =& \, \frac{i f_V^2 h_P}{4} \int \df x \te{d}^4\! y \int \frac{\te{d}^4\! k \te{d}^4\! p}{(2\pi)^8} \, e^{i (k-p)(x-y)} \bra P [\Dvec^0(k)]^{-1}_{\mu\nu\alpha\beta} \right. \\
		& \left. \cdot \zeta^{\alpha\beta}(y) [\Dch^0(p)]^{-1} P_1^{\mu\nu\rho\sigma} g_{\rho\eta} (-i k^\eta - \partial^\eta ) \vartheta_\sigma(x) \ket.
\end{align*}
First, the term proportional to $k^\eta$ is calculated. For that, the integral is localised, \textit{i.e.}, $y =: x-z$ and $\zeta(y)$ is expanded at $z {=}0$ up to $\order(Q^4)$, 
\begin{align*}
	I_1 = \frac{f_V^2 h_P}{4} & \int \df x \te{d}^4\! z \int \frac{\te{d}^4\! k \te{d}^4\! p}{(2\pi)^8} \, e^{i (k-p)z} k_\eta \bra P [\Dvec^0(k)]^{-1}_{\mu\nu\alpha\beta} \right. \\
		& \hspace{-2em} \left. \cdot ( 1 - z_\tau \partial^\tau_z) \zeta^{\alpha\beta}(z) [\Dch^0(p)]^{-1} P_1^{\mu\nu\rho\sigma} g_{\rho\eta} \vartheta_\sigma(x) \ket \\
		& \hspace{-2.75em} + \order(Q^6).
\end{align*}
Thereby, the first term in the expansion of $\zeta$ yields zero due to the odd number of $k$'s. For the second term, the integration over $\te{d}^4 \! z$ can be performed after partial integration with respect to $z$ as described in subsection \ref{subsec:vec},
\begin{align*}
	-z_\tau e^{i(k-p)z} [\Dch^0(p)]^{-1} \ \rightarrow \ i e^{i(k-p)z} \partial_\tau^p [\Dch^0(p)]^{-1} ,
\end{align*}
yielding the $\delta$-function $\delta(k-p)$ in momentum space. So, the resulting integral over one space coordinate $x$ and one momentum coordinate $k$ is given by
\begin{align*}
	I_1 =& \, - \frac{i f_V^2 h_P}{2} \int \df x \int \frac{\te{d}^4\! k}{(2\pi)^4} \bra P [\Dvec^0(k)]^{-1}_{\mu\nu\alpha\beta} \, \partial^\tau_x \zeta^{\alpha\beta}(x) \right. \\
		& \hspace{2.5em} \left. \cdot \, [\Dch^0(k)]^{-2} \, P_1^{\mu\nu\rho\sigma} g_{\rho\eta} \vartheta_\sigma(x) \ket  k_\eta k_\tau  + \order(Q^6) \adb \\
		=& \, - \frac{i f_V^2 h_P}{4 F^2} \int \df x \int \frac{\te{d}^4\! k}{(2\pi)^4} \, k_\eta k_\tau \,(k^2 - \mv^2)^{-1} P_{V \! , \, \mu\nu\alpha\beta} \\
		& \hspace{0.5em} \cdot \, \partial^\tau_x \zeta^{\alpha\beta}_{aP}(x) \, (k^2 - \Mp^2)^{-2} \, P_1^{\mu\nu\rho\sigma} g_{\rho\eta} \vartheta_\sigma^{Pa}(x) + \order(Q^6) 
\end{align*}
since the part proportional to $P_{A, \, \mu\nu\alpha\beta}$ yields zero. The two propagators can be reformulated using Feynman parameters,
\begin{align*}
	&\left[ (k^2 - \mv^2) (k^2 - \Mp^2)^2 \right]^{-1} \\
	&= \frac{\Gamma(3)}{\Gamma(1) \Gamma(2)} \int_0^1 \df u \, \frac{u}{\left[(1-u)(k^2-\mv^2) + u(k^2- \Mp^2)\right]^3} \nn \\
	&= 2 \int_0^1 \df u \, \frac{u}{\left[k^2 - M(u)\right]^3} \, , \adb \\
	& M(u) := (1-u) \mv^2 + u \Mp^2 \,.
\end{align*}
Furthermore, the momentum vectors contracted with the corresponding Lorentz indices can be substituted according to Eq.\ \eqref{eq:Repl-kmu} as
\begin{align*}
	&k_\eta k_\tau P_{V,\mu\nu\alpha\beta} P_1^{\mu\nu\rho\sigma} g_{\rho \eta} = \frac{1}{2} k_\tau \left( k_\alpha g_{\beta \rho} - k_\beta g_{\alpha\rho} \right) g^{\rho\sigma} \\
	\rightarrow & \, \frac{1}{4+2 \varepsilon} \, k^2 P_{1,\alpha\beta \tau \rho} \, g^{\rho\sigma}.
\end{align*}
With this substitution the momentum integral can be determined as (cf.\ Eq.\ \eqref{eq:FormularMomInt})
\begin{align*}
	\int \frac{\te{d}^4 \! k}{(2 \pi)^4} \, \frac{k^2}{\left[k^2 - M(u)\right]^3}
	&= \frac{i}{16 \pi^2} \left( \frac{m^2}{4 \pi \mu^2} \right)^{\!\! \varepsilon} \frac{ \Gamma(3 + \varepsilon) \Gamma(- \varepsilon) }{\Gamma(3) \Gamma(2+ \varepsilon)} \\
	&= - i \bar{\lambda} + (\te{finite}).
\end{align*}
So, the integral $I_1$ is given as
\begin{align*}
	I_1 &= - \frac{f_V^2 h_P}{8 F^2} \int \df x \,  P_{1,\alpha\beta \tau \rho} \, g^{\rho\sigma} \, \partial^\tau_z \zeta^{\alpha\beta}_{aP}(x) \, \vartheta^{Pa}_\sigma(x) \nn \\
	& \phantom{=} \ + (\te{finite}) + \order(Q^6) \\
	&= \frac{f_V^2 h_P}{8 F^2} \int \df x \,  \bra \zeta^{\tau \eta}(x) \, \partial^\tau \vartheta_\eta(x) \ket + (\te{finite})+ \order(Q^6) .
\end{align*}
The second part of the integral $I$ can be calculated in a similar way yielding
\begin{align*}
	I &= I_1 + I_2 = I_1 + \{ - I_1 + (\te{finite}) + \order(Q^6) \} \\
	 &= (\te{finite}) + \order(Q^6).
\end{align*}

\end{appendix}

\bibliography{literature-paperInf}

\end{document}